\documentclass[12pt,preprint]{aastex}
\usepackage{array}
\usepackage{longtable}
\usepackage{epsfig}
\usepackage{natbib}
\usepackage{graphicx}
\usepackage{tabularx}
\usepackage{rotating}
\usepackage{multirow}
\usepackage{lscape}
\usepackage{mathrsfs,amssymb}
\usepackage{amsmath}
\usepackage{subfigure}
\usepackage{ulem}
\usepackage{color}
\usepackage{rotating}

\newcommand    \mum      {\,{\rm \mu m}}
\newcommand    \Rv         {R_{V}}
\newcommand    \AKs       {A_{\rm K_s}}
\newcommand    \Ks         {K_{\rm s}}

\shorttitle{Four percent distance precision to the Galactic Center}
\shortauthors{X. D. Chen et al.}

\begin{document}

\title{An extremely low mid-infrared extinction law toward the
  Galactic Center and 4\% distance precision to 55 classical Cepheids}

\author{Xiaodian Chen\altaffilmark{1}, 
Shu Wang\altaffilmark{2}, 
Licai Deng\altaffilmark{1}, and 
Richard de Grijs\altaffilmark{2,3,4}
}
\altaffiltext{1}{Key Laboratory for Optical Astronomy, National
  Astronomical Observatories, Chinese Academy of Sciences, 20A Datun
  Road, Chaoyang District, Beijing 100012, China;
  chenxiaodian@nao.cas.cn}
\altaffiltext{2}{Kavli Institute for Astronomy \& Astrophysics, Peking
  University, Yi He Yuan Lu 5, Hai Dian District, Beijing 100871,
  China; shuwang@pku.edu.cn}
\altaffiltext{3}{International Space Science Institute--Beijing, 1
  Nanertiao, Zhongguancun, Hai Dian District, Beijing 100190, China}
\altaffiltext{4}{Department of Physics and Astronomy, Macquarie
  University, Balaclava Road, North Ryde, NSW 2109, Australia}

\begin{abstract}
Distances and extinction values are usually degenerate. To refine the
distance to the general Galactic Center region, a carefully determined
extinction law (taking into account the prevailing systematic errors)
is urgently needed. We collected data for 55 classical Cepheids
projected toward the Galactic Center region to derive the near- to
mid-infrared extinction law using three different approaches. The
relative extinction values obtained are $A_J/A_{\Ks}=3.005,
A_H/A_{\Ks}=1.717, A_{[3.6]}/A_{\Ks}=0.478, A_{[4.5]}/A_{\Ks}=0.341,
A_{[5.8]}/A_{\Ks}=0.234, A_{[8.0]}/A_{\Ks}=0.321,
A_{W1}/A_{\Ks}=0.506$, and $A_{W2}/A_{\Ks}=0.340$. We also calculated
the corresponding systematic errors. Compared with previous work, we
report an extremely low and steep mid-infrared extinction law. Using a
seven-passband `optimal distance' method, we improve the mean distance
precision to our sample of 55 Cepheids to 4\%. Based on four confirmed
Galactic Center Cepheids, a solar Galactocentric distance of
$R_0=8.10\pm0.19\pm0.22$ kpc is determined, featuring an uncertainty
that is close to the limiting distance accuracy (2.8\%) for Galactic
Center Cepheids.
\end{abstract}

\keywords{ISM: dust, extinction --- distance scale --- Galaxy: center
  --- Galaxy: bulge --- stars: variables: Cepheids --- infrared: star}

\section{Introduction}

The Galactic bulge is a complex environment. It is affected by heavy
extinction, which prevents us to some extent from undertaking detailed
studies of its stellar populations. Until recently, the Galactic bulge
was thought to only contain very old stars, but we now know of the
presence of a 200 pc nuclear stellar disk in the Galactic Center
\citep{Serabyn96, vanLoon03, Matsunaga11}. This nuclear disk is
composed of young stars with ages ranging from a few million to a
billion years. The distance to the Galactic Center is a fiducial
distance, adoption of which will affect calculations of the Galaxy's
mass, luminosity, and the rotation speed at the solar circle
\citep[e.g.,][]{deGrijs17}. To independently refine this distance and
better trace the bulge's structure, classical Cepheids can be used as
important and accurate stellar distance tracers.

Since the first three genuine Galactic Center Cepheids were found by
\citet{Matsunaga11}, an additional 52 Cepheids have been found along
this general sightline \citep{Dekany15a, Dekany15b,
  Matsunaga16}. Analysis of these variables has the potential to
uncover some of the remaining secrets of the Galactic bulge. However,
the different near-infrared (NIR) extinction laws commonly adopted
introduce systematic distance uncertainties of at least 10\%
\citep{Matsunaga16}. \citet{Dekany15b} found a dozen Cepheids in the
bulge, a conclusion based on their distance estimates of around 9.5
kpc. \citet{Matsunaga16} supported a larger distance to these
variables; they also ruled out the presence of any other Cepheids in
the bulge in addition to those in the nuclear stellar disk. To
properly distinguish between both claims, determination of an unbiased
NIR extinction law, including a detailed assessment of the systematic
errors, is required.

Usually, NIR and mid-infrared (MIR) extinction laws based on stellar
samples are determined in one of two ways. One method is based on
analysis of the color excess diagram, specifically in the regime
populated by red giants (RGs) and red clump (RC) stars. The second
approach uses the color excess--extinction diagram of RC stars that
are all located at similar distances. The former method is convenient,
but its application will introduce large systematic errors since the
slope of the color excess--color excess diagram is sensitive to
contamination and distance differences. Another concern relates to the
use of the NIR power-law $A_{\lambda}\propto\lambda^{-\alpha}$
hypothesis, which will introduce a large systematic bias when using
$E(J-K_{\rm s})/E(H-K_{\rm s})$ to derive $\alpha$. The latter method
can only be applied explicitly to objects residing in a spatially
tightly confined volume, such as those in the Galactic bulge. The
associated systematic errors mainly come from the scatter in the
objects' distances, as well as in the absolute magnitudes and
intrinsic colors of the RC stars. Based on careful sample selection,
the systematic error associated with this latter method can be much
lower than that affecting the former method. However, given that we
have to adopt a number of important assumptions, the prevailing
systematic error is not easily calculated. This is, hence, currently
an open issue.
  
Compared with RGs and RC stars, Cepheids are better suited for
distance analysis given their well-studied physical properties, tight
period--luminosity relations (PLRs), and high luminosities. The tight
PLRs allow us to use Cepheids to determine the extinction law. In
addition, by employing Cepheids the resulting systematic error should
be among the lowest attainable for extinction law
determinations. Based on our sample of some 50 Cepheids, the
associated statistical error is indeed much smaller than the
prevailing systematic error (see Section 3.1.2). This means that
Cepheids are ideal objects not only to anchor the distance scale, but
also to constrain the extinction law. To reduce the uncertainties in
Cepheid distances, we also need to constrain the scatter in the
PLRs. The 1$\sigma$ dispersion in the NIR PLRs of \citet{Chen17} is
approximately 0.10 mag, which amounts to a 5\% uncertainty in the
distance scale. To reduce this uncertainty, independent distances or
multi-passband PLRs are needed.

In this paper, we have collected a sample of classical Cepheids along
lines of sight toward the Galactic Center for which NIR and MIR data
are available (see Section 2). Using these Cepheids, we derive the
NIR--MIR extinction law using both the color excess--extinction method
and the simple color excess method (see Section 3). More importantly,
we discuss the errors in the resulting extinction law in
detail. Combining the NIR and MIR bands, in Section 4 we achieve an
improved distance precision of 4\% for our Cepheid sample. An overview
of the Cepheid distribution and derivation of the Galactic Center
distance are provided in Section 4 as well. A discussion of the
diversity of NIR extinction laws, derivation of an extremely low MIR
extinction law, and an overview of the absolute extinction in the
Galactic Center are covered in Section 5. We conclude the paper in
Section 6.

\section{Data}

We collected our sample of classical Cepheids from
\citet{Matsunaga11,Matsunaga16} and \citet{Dekany15a,
  Dekany15b}. \citet{Matsunaga11} provide details about three Cepheids
in the Galactic Center's nuclear disk, \citet{Dekany15a} found a pair
of Cepheids, \citet{Dekany15b} list 35 Cepheids located in the
Galactic longitude range $-10.5^\circ \lesssim l \lesssim +10^\circ$,
and \citet{Matsunaga16} provide a compilation of 29 Cepheids in the
general Galactic Center region. After excluding duplicate sources, the
total number of Cepheids left is 55.

The NIR data are taken from these articles. \citet{Matsunaga11,
  Matsunaga16} used the Infrared Survey Facility (IRSF) 1.4 m
telescope equipped with the SIRIUS camera to observe their sample
objects through the $J$, $H$, and $K_{\rm s}$ NIR filters. The
effective wavelengths of the SIRIUS filters are slightly different
from those of the Two Micron All-Sky Survey (2MASS) $J$, $H$, and
$K_{\rm s}$ bands; the SIRIUS zero-point calibration is discussed by
\citet{Nishiyama06}. The data of \citet{Dekany15a, Dekany15b}
originate from the VISTA Variables in the V\'ia L\'actea (VVV)
survey. The effective wavelengths of the VVV $J$, $H$, and $K_{\rm s}$
filters are 1.254 $\mu$m, 1.646 $\mu$m, and 2.149 $\mu$m, respectively
\citep{Saito12}; the magnitude differences with respect to 2MASS are
discussed by \citet{Gonzalez11}. The VVV's mean magnitudes in the $J$
and $H$ bands are determined based on one to five epochs of
observations, while the $K_{\rm s}$ band has more than 30 epochs of
data. Since the VVV data are composed of larger numbers of
observations, we convert the IRSF magnitudes to VVV magnitudes using
the transformation equations of \citet{Nishiyama06} and
\citet{Gonzalez11}.

We also collected MIR data from the {\sl Spitzer Space Telescope}'s
Galactic Legacy Infrared Mid-Plane Survey Extraordinaire (GLIMPSE) II
and the Wide-Field Infrared Survey Explorer (WISE) survey for our 55
sample Cepheids. GLIMPSE II used four Infrared Array Camera (IRAC)
bands, referred to as [3.6], [4.5], [5.8], and [8.0], which are
characterized by isophotal wavelengths of 3.55, 4.49, 5.73, and 7.87
$\mu$m, respectively. The survey covers the area where our 55 Cepheids
are located. Since the GLIMPSE II source catalog provides the
corresponding 2MASS $K_{\rm s}$ photometry, all of our Cepheids are
carefully matched based on both their positions and their $K_{\rm s}$
magnitudes. The limiting and saturation magnitudes in the [3.6],
[4.5], [5.8], and [8.0] filters are, respectively, 15.5, 15.0, 13.0,
and 13.0 mag (limiting magnitudes), and 7.0, 6.5, 4.0, and 4.0 mag
(saturation magnitudes) \citep{Churchwell09}. All Cepheid luminosities
are within the detection range. The GLIMPSE II survey is characterized
by two-epoch coverage, each with three visits on the sky; each Cepheid
has 3--6 detections. The mean magnitude is weighted by the individual
photometric errors.
WISE is a full-sky survey undertaken in four bands: $W1$ (3.35
$\mu$m), $W2$ (4.60 $\mu$m), $W3$ (11.56 $\mu$m), and $W4$ (22.09
$\mu$m). Its 5$\sigma$ detection limits are 16.5, 15.5, 11.2, and 7.9
mag, respectively \citep{Wright10}. For our Cepheid sample, the $W4$
magnitudes are below the limiting magnitude. Although the $W3$
magnitudes lie within the detection range, few sources have reliable
$W3$ magnitudes because of the dense and crowded stellar environment
in the bulge. Therefore, we only consider the Cepheids' $W1$ and $W2$
photometry. 

WISE is a multi-epoch survey; the weighted average magnitude is
estimated based on the photometric errors associated with the
individual exposures. In crowded regions, the survey's relatively
  low spatial resolution may significantly affect the quality of the
  resulting photometry. Note that the spatial resolution of WISE is
  around 6 arcsec \citep{Wright10} which is much worse than that of
  GLIMPSE II, which is around 1.2 arcsec \citep{Churchwell09}. For
  this reason, approximately half of the known Galactic Center
  Cepheids are missing in the WISE catalog, while the other half have
  large photometric uncertainties. Upon checking the appearance of the
  Cepheids in the GLIMPSE II mosaic image, we concluded that the
  nuclear star cluster is too crowded for individual stars to be
  easily distinguished at the prevailing {\sl Spitzer}
  resolution. However, our Cepheid sample objects are located well
  outside the nuclear star cluster. In addition, the Cepheids are very bright
  ([3.6] $\lesssim 11$ mag and $[3.6] \lesssim 9$ mag for all Cepheids
  and for the central Cepheids, respectively), which implies that the
  effects of crowding for these stars are much smaller than for
  fainter objects. All photometric information is summarized in Table
\ref{t1}. The four Cepheids located in the inner Galactic Center
  region itself are also indicated.

\section{NIR--MIR Extinction Laws along Sightlines to the Galactic Center Region}

In order to use these Cepheids to determine the NIR--MIR extinction
law, we first need to calculate their absolute magnitudes and
intrinsic colors. Since classical Cepheids occupy a narrow instability
strip and follow tight PLRs, the systematic uncertainties in the
absolute magnitudes and intrinsic colors are expected to be smaller
than for other tracers like RC stars and RGs. \citet{Chen17} derived
NIR PLRs for Galactic classical Cepheids based on direct distance
estimates obtained by application of the open cluster main-sequence
fitting method. \citet{Wang18} derived PLRs in eight MIR filters based
on a large sample of Galactic classical Cepheids, with indirect
distances determined using the NIR `optimal distance'
method. Therefore, we calculated the mean absolute magnitudes for all
Cepheids in nine bands, $M_\lambda$ ($\lambda: J, H, \Ks$, [3.6],
[4.5], [5.8], [8.0], $W1$, and $W2$). The intrinsic colors for any
pair of bands are thus available, e.g., $(\Ks-\lambda)_0 = M_{\Ks} -
M_\lambda$. In this paper, we use the $\Ks$ band as our reference band
to determine the extinction, since the $\Ks$ band has the smallest
photometric error and is usually used as the basis for NIR distances
and extinction determinations.

We adopted three methods to determine the extinction along Galactic
Center sightlines, including (1) the color excess--extinction method;
(2) the simple color excess method; and (3) assessment of the absolute
extinction for the four Cepheids in the Galactic Center.
If the extinction curves along these sightlines are found to exhibit
variation, this could also help us to determine whether that variation
might be caused by employing the different methods or by environmental
variations along the different sightlines.

\begin{sidewaystable}
\begin{tiny}
\begin{center}
\hspace{0.0in}

\begin{longtable}{ccccccccccccc}
\caption{NIR and MIR mean magnitudes of 55 classical Cepheids in the Galactic bulge.}\label{t1}  \\
  \hline
  R.A. (J2000) & Dec. (J2000)  & Period & $\langle J \rangle$ & $\langle H \rangle$ & $\langle K_{\rm s} \rangle$ & [3.6]  & [4.5]  & [5.8]  & [8.0]   & $W1$ & $W2$  & Center Cepheid    \\
(hh:mm:ss)  & $(^\circ$ $'$ $''$)  & (days)    & (mag)  & (mag)    & (mag)              & (mag)& (mag)& (mag) & (mag) &(mag)  &  (mag)   & y    \\
  \hline
18:01:24.49 & $-$22:54:44.6 & 11.2397 & 19.13(20) & 14.76(05) & 12.71(02) & 11.164(054) & 10.749(076) & 10.527(052) & 10.714(067) & 11.230(053) & 10.987(059)&\\    
18:01:25.09 & $-$22:54:28.3 & 11.2156 & 18.41(12) & 14.67(05) & 12.66(02) & 11.187(035) & 10.737(054) & 10.467(051) & 10.599(067) &             &            &\\    
17:22:23.24 & $-$36:21:41.5 & 7.5685  & 15.69(03) & 13.32(07) & 12.01(02) & 10.961(077) & 10.625(053) & 10.328(091) & 10.455(059) &             &            &\\    
17:21:16.05 & $-$36:43:25.2 & 6.3387  & 17.22(07) & 14.19(10) & 12.57(02) & 11.256(056) & 10.872(051) & 10.724(061) & 10.790(069) & 11.294(238) & 10.810(129)&\\    
17:20:14.62 & $-$37:11:16.0 & 5.0999  & 15.25(04) & 13.20(08) & 12.04(02) & 11.287(038) & 11.088(070) & 10.966(059) & 11.057(086) & 11.157(057) & 11.069(067)&\\    
17:26:34.72 & $-$35:16:24.1 & 13.4046 & 16.94(09) & 13.51(05) & 11.70(02) & 10.436(059) & 10.197(055) &  9.890(052) & 10.011(059) &             &            &\\    
17:25:29.70 & $-$34:45:45.9 & 12.3267 & 16.87(07) & 13.61(04) & 11.86(02) & 10.488(027) & 10.272(049) & 10.024(038) & 10.125(066) & 10.450(034) & 10.109(034)&\\    
17:26:43.41 & $-$34:58:25.6 & 9.8383  & 	        & 16.66(11) & 13.86(02) & 11.903(049) & 11.413(053) & 11.040(076) & 11.482(141) &             &            &\\    
17:26:54.24 & $-$35:01:08.2 & 4.2904  &           & 15.94(05) & 13.93(02) & 12.418(064) & 11.989(068) & 11.535(090) &             &             &            &\\    
17:30:46.64 & $-$34:09:04.4 & 4.7272  &           & 15.36(06) & 13.41(02) & 12.026(039) & 11.714(063) & 11.474(090) & 11.792(303) &             &            &\\    
17:28:15.86 & $-$34:32:27.2 & 7.0235  &           & 15.62(06) & 13.27(02) & 11.494(054) & 11.026(049) & 10.782(070) & 10.914(059) &             &            &\\    
17:38:42.96 & $-$31:44:55.7 & 11.9663 & 16.39(09) & 13.78(04) & 12.01(03) & 10.633(041) & 10.287(065) & 10.120(051) & 10.212(051) & 10.804(066) & 10.509(075)&\\    
17:36:44.46 & $-$32:04:38.6 & 8.3220  &           & 15.83(07) & 13.09(03) & 11.342(049) & 10.936(053) & 10.670(065) & 10.938(084) &             &            &\\    
17:40:41.72 & $-$30:48:46.9 & 10.6634 & 17.39(08) & 13.96(07) & 12.13(04) & 10.981(063) & 10.546(072) & 10.450(065) & 10.592(067) &             &            &\\    
17:40:25.15 & $-$31:04:50.5 & 4.7004  & 16.04(05) & 13.48(06) & 12.10(04) & 11.040(040) & 10.793(057) & 10.626(068) & 10.994(074) &             &            &\\    
17:42:20.00 & $-$30:14:50.7 & 23.9729 &           & 14.66(07) & 12.05(04) & 10.063(062) &  9.546(051) &  9.328(040) &  9.493(042) & 10.553(064) &  9.961(060)&\\    
17:41:15.13 & $-$30:07:17.7 & 13.3262 & 16.88(14) & 13.41(10) & 11.62(04) & 10.376(091) &  9.973(057) &  9.799(077) & 10.126(058) &             &            &\\    
17:51:05.72 & $-$26:38:18.3 & 6.3496  & 18.11(07) & 14.69(05) & 12.85(03) & 11.478(050) & 11.139(058) & 10.958(066) & 11.023(178) &             &            &\\    
17:51:13.77 & $-$26:48:55.9 & 12.9488 &           & 14.08(06) & 12.13(03) & 10.693(077) & 10.324(064) & 10.098(055) & 10.171(050) & 10.738(063) & 10.265(041)&\\    
17:49:41.42 & $-$27:27:14.6 & 10.4762 & 15.51(04) & 12.91(06) & 11.48(03) & 10.203(070) & 10.063(062) & 10.174(046) &             &             &            &\\    
17:50:30.49 & $-$27:13:46.7 & 12.6433 &           & 14.62(05) & 12.30(03) & 10.779(043) & 10.403(058) & 10.138(042) & 10.432(044) &             &            &\\    
17:53:16.07 & $-$26:28:26.9 & 5.9995  &           & 15.33(05) & 13.08(03) & 12.341(038) & 11.559(047) & 11.284(080) &             &             &            &\\    
17:52:21.66 & $-$26:31:19.3 & 11.9921 & 17.54(08) & 14.27(06) & 12.25(03) & 10.711(046) & 10.395(059) & 10.125(053) & 10.128(039) &             &            &\\    
17:51:43.80 & $-$26:31:11.3 & 5.3407  &           & 15.47(05) & 13.25(03) & 12.007(064) & 11.500(076) & 11.119(130) &             &             &            &\\    
17:55:44.68 & $-$25:00:30.2 & 6.7911  & 17.65(10) & 14.37(07) & 12.64(03) & 11.389(067) & 11.133(052) & 10.880(079) &             &             &            &\\    
17:58:26.68 & $-$23:52:08.3 & 4.5517  & 18.81(22) & 15.31(07) & 13.37(02) & 11.990(073) & 11.563(092) & 11.145(082) & 11.525(177) &             &            &\\    
18:03:31.12 & $-$22:21:14.0 & 10.9622 &           & 14.67(04) & 12.45(02) & 10.971(039) & 10.621(045) & 10.315(053) & 10.523(060) & 11.028(072) & 10.339(049)&\\    
18:09:14.03 & $-$20:03:21.4 & 24.8683 &           & 16.52(04) & 13.00(02) & 10.477(063) &  9.892(059) &  9.372(047) &  9.519(045) & 10.904(038) &  9.993(033)&\\    
17:22:10.10 & $-$36:44:18.8 & 14.8719 & 16.63(05) & 13.19(09) & 11.44(02) & 10.149(089) & 10.019(049) &  9.670(050) &  9.798(042) & 10.174(035) &  9.759(031)&\\    
17:26:00.10 & $-$35:15:15.0 & 18.2396 & 16.30(10) & 12.91(04) & 11.09(02) &  9.690(026) &  9.346(047) &  9.140(039) &  9.247(033) &  9.789(030) &  9.411(028)&\\    
17:32:14.07 & $-$33:23:59.5 & 9.9061  & 	        & 12.09(07) & 11.01(02) & 10.276(041) & 10.265(076) & 10.063(045) & 10.066(041) &  9.611(030) &  9.533(029)&\\    
17:40:51.51 & $-$30:24:53.2 & 7.9311  & 15.42(06) & 13.05(09) & 11.72(04) & 10.640(049) & 10.505(054) & 10.370(054) & 10.513(059) &             &            &\\    
17:40:24.58 & $-$31:01:32.9 & 17.3711 & 15.80(09) & 12.77(06) & 11.18(04) &  9.793(034) &  9.580(040) &  9.383(037) &  9.431(033) &             &            &\\    
17:50:17.54 & $-$27:08:13.3 & 7.0503  & 16.79(06) & 13.90(06) & 12.35(03) & 11.296(108) & 11.047(057) & 10.879(054) & 11.063(058) &             &            &\\    
17:55:24.20 & $-$25:30:22.3 & 15.4472 & 16.60(05) & 13.16(06) & 11.46(03) & 10.313(047) &  9.951(052) &  9.669(049) &  9.980(049) &             &            &\\    
17:54:40.25 & $-$25:34:39.5 & 17.1620 & 17.33(09) & 13.47(06) & 11.47(03) & 10.118(045) &  9.798(052) &  9.411(106) &             & 10.135(032) &  9.359(025)&\\    
17:56:01.96 & $-$25:15:44.9 & 13.4540 & 16.01(04) & 13.03(07) & 11.42(03) & 10.259(043) &  9.964(048) &  9.778(044) &  9.904(051) &             &            &\\    
17:20:18.26 & $-$36:58:52.8 & 9.944	  & 15.22(02) & 12.83(02) & 11.54(02) & 10.660(068) & 10.594(051) & 10.273(051) & 10.558(060) & 10.956(072) & 10.414(041)&\\    
17:24:12.58 & $-$36:01:46.9 & 17.597	& 13.66(02) & 11.58(02) & 10.46(02) &  9.555(028) &  9.520(042) &  9.263(031) &  9.233(036) &  9.743(030) &  9.742(051)&\\    
17:29:59.17 & $-$34:09:55.1 & 31.52	  & 12.30(02) & 10.30(02) &  9.20(02) &  8.424(035) &  8.403(042) &  8.154(026) &  8.135(025) &  9.116(036) &  8.184(018)&\\    
17:38:46.14 & $-$31:26:22.8 & 25.03	  &           & 12.31(02) & 10.77(02) &  9.365(040) &  9.138(041) &  8.926(030) &  8.974(030) &  9.592(025) &  9.003(024)&\\    
17:40:41.03 & $-$30:41:38.6 & 23.88	  & 15.16(02) & 12.08(02) & 10.47(02) &  9.361(032) &  9.029(043) &  8.864(027) &  9.014(033) &  9.529(031) &  9.166(029)&\\    
17:44:56.91 & $-$29:13:33.8 & 18.87	  & 14.95(20) & 12.14(02) & 10.32(02) &  9.060(065) &  8.650(109) &  8.520(078) &             &     	      &            &y\\    
17:50:11.26 & $-$27:19:42.9 & 16.52	  &           & 12.58(02) & 10.95(02) & 10.087(049) &  9.824(042) &  9.529(045) &  9.621(044) & 10.283(055) &  9.566(044)&\\    
17:52:28.94 & $-$26:23:40.0 & 22.63	  & 15.55(02) & 12.47(02) & 10.91(02) &  9.661(046) &  9.495(035) &  9.196(039) &  9.303(026) & 10.073(037) &  9.685(048)&\\    
17:52:38.14 & $-$26:19:43.3 & 38.16	  & 14.01(02) & 11.22(02) &  9.86(02) &  8.689(077) &  8.583(068) &  8.336(036) &  8.323(030) &  8.952(025) &  8.689(030)&\\    
17:57:31.41 & $-$24:30:26.7 & 24.32	  &           & 13.81(02) & 11.74(02) &  9.832(036) &  9.620(067) &  9.303(038) &  9.386(035) &             &            &\\    
18:02:14.84 & $-$22:27:10.7 & 17.12	  &           & 13.90(02) & 11.99(02) & 10.488(031) & 10.167(047) & 10.001(041) &  9.969(050) &             &            &\\    
18:03:29.93 & $-$22:03:22.5 & 39.68	  & 14.55(02) & 11.68(02) & 10.35(02) &  9.123(042) &  8.932(045) &  8.725(030) &  8.747(027) &  9.382(027) &  8.979(023)&\\    
18:03:53.95 & $-$21:58:11.7 & 45.3	  & 15.42(20) & 11.91(02) & 10.23(02) &  8.939(048) &  8.810(051) &  8.513(031) &  8.605(031) &  8.347(021) &  8.159(030)&\\    
18:05:52.84 & $-$21:06:41.9 & 21.41	  &           & 13.10(02) & 11.44(02) & 10.085(028) &  9.873(045) &  9.573(030) &  9.688(030) & 10.432(042) & 10.129(054)&\\    
18:07:07.82 & $-$20:34:50.1 & 19	    &           & 12.74(02) & 10.94(02) &  9.787(041) &  9.522(043) &  9.276(033) &  9.382(042) &  9.828(029) &  9.322(022)&\\    
17:46:06.01 & $-$28:46:55.1& 23.5380 & 15.63(02) & 12.04(02) & 10.18(02) &  8.840(033) &  8.540(045) &  8.240(038) &  8.240(188) &  8.887(130) &  8.637(110) & y\\    
17:45:32.27 & $-$29:02:55.2& 19.9600 & 15.42(02) & 12.00(02) & 10.17(02) &  &   &   &   &             &          &  y\\                                        
17:45:30.89 & $-$29:03:10.5 & 22.7600 & 16.36(02) & 12.44(02) & 10.35(02) &  8.880(044) &  8.530(051) &  8.350(035) &  8.730(118) &             &            & y\\ 
  \hline
\end{longtable}

\end{center} 

\end{tiny}
\end{sidewaystable}

\clearpage

\subsection{The Color Excess--Extinction Method} 
\subsubsection{Distance Modulus and Relative Extinction}

\begin{figure}[h!]
\centering
\vspace{-0.0in}
\includegraphics[angle=0,width=160mm]{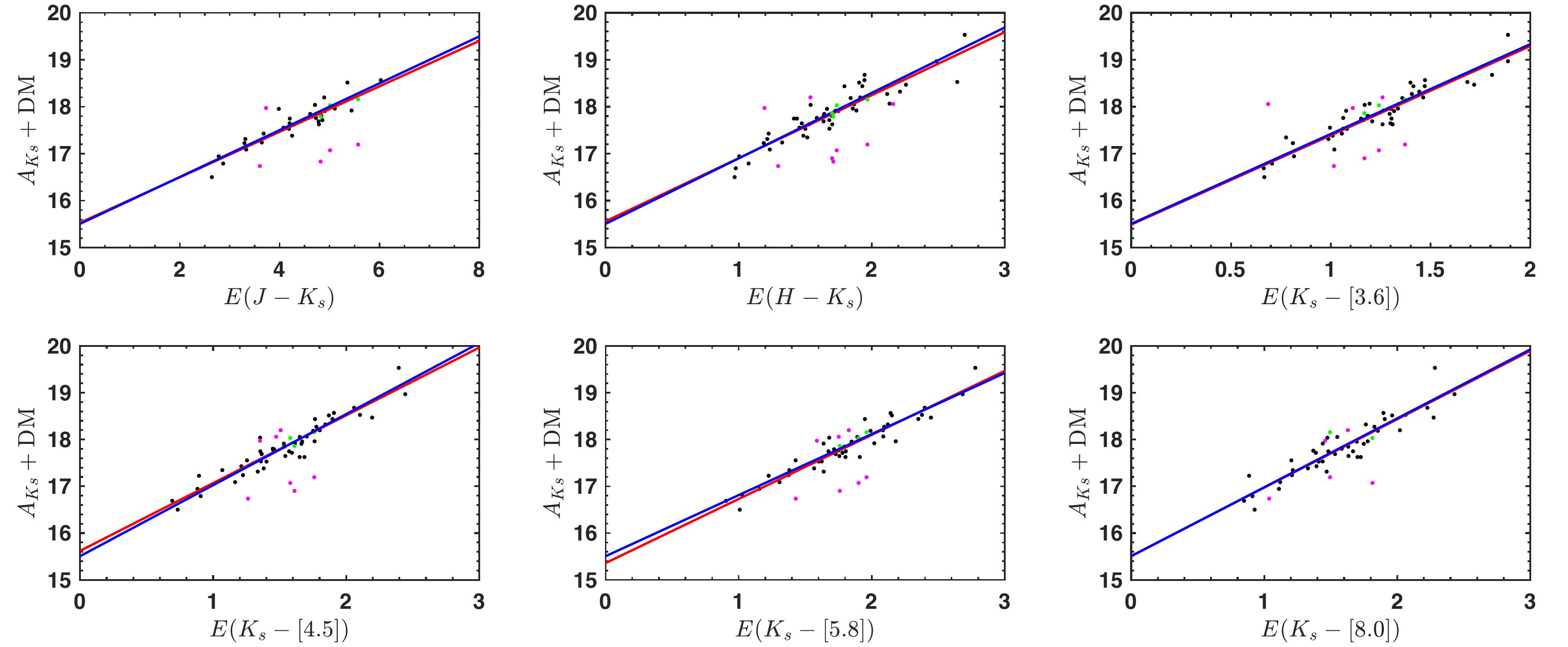}
\vspace{-0.0in}
\caption{\label{f1} Color excess--extinction diagrams for our 55
  Cepheids. The black dots represent the 46 Cepheids adopted for
    our fits. The magenta dots are the nine Cepheids that were
    excluded, including the four inner Galactic Center Cepheids; the
    latter are shown in green at their adjusted distances (distance
    moduli adjusted by +0.96 mag).} The red lines are the best linear
  fits without having adopted a specific DM, while the blue lines are
  the best linear fits for DM = 15.506 mag.
\end{figure}

We first use the color excess vs. extinction diagram to determine the
extinction. Here, the $x$ axis is the color excess, e.g., $E(J-\Ks)$,
and the $y$ axis is the absolute extinction plus the relevant distance
modulus (DM), e.g., $A_{\Ks}+{\rm DM}=\langle K\rangle-\langle M_{Ks}
\rangle$. Therefore, the slope of this diagram represents the
extinction vector. If a source has no color excess and zero
extinction, it coincides with the diagram's intercept, which
corresponds to the DM. We refer to this method as the color
excess--extinction method. To adopt this method to determine the
extinction, the Cepheids need to be located at similar distances.
Since nine Cepheids in our sample have distances that fall outside the
2$\sigma$ uncertainty envelope around our best linear fits, we used a
sigma-clipping approach to exclude them. In Fig. \ref{f1}, the black
dots are the remaining 46 Cepheids located in the Galactic plane
($-0.1^\circ \lesssim b \lesssim +0.1^\circ$) and spread across the
Galactic longitude range $-10.5^\circ \lesssim l \lesssim +10^\circ$
(except for the central two degrees, $-1^\circ \lesssim l \lesssim
+1^\circ$). We combine six color excesses in the different bands ($J,
H$, [3.6], [4.5], [5.8], and [8.0]) with respect to the $K_{\rm s}$
band to precisely determine the DM. The red lines in Fig. \ref{f1}
show the best linear fits to the 46 Cepheids; their slopes and
intercepts represent the extinction law and mean DM. Combining these
six intercepts, we derive a mean DM = $15.506\pm0.108$ mag for these
46 Cepheids. Therefore, we fix the intercept to 15.506 mag and
subsequently rederive the slopes: see the blue lines. The best-fitting
parameters are listed in Table \ref{t2}. Both sets of results are
comparable within the 1$\sigma$ statistical errors.

\begin{table}[h!]
\begin{center}
\caption{\label{t2}Linear fit parameters from Fig. \ref{f1}.}
\vspace{0.15in}
\begin{tabular}{lccc}
\hline
\hline
   &\multicolumn{2}{c|}{$A_{\Ks}+{\rm DM}=a \times E(\lambda-K_{\rm s})+b$} & $A_{\Ks}=c\times E(\lambda-K_{\rm s}), {\rm DM}=15.506$ mag \\ 
 \hline 
$\lambda$ &        $a$     &      $b$       &       $c$\\
 \hline         
$J$  &   $0.484 \pm0.043$ & $15.53\pm0.19$ & $0.4988 \pm0.008$\\  
$H$ &   $1.343 \pm0.092$ & $15.56\pm0.16$ & $1.394 \pm0.021$\\
$[3.6]$ &   $-1.901\pm0.111$ & $15.49\pm0.14$ & $-1.914\pm0.025$\\  
$[4.5]$ &   $-1.448\pm0.076$ & $15.62\pm0.12$ & $-1.518\pm0.019$\\  
$[5.8]$ &  $-1.369\pm0.067$ & $15.36\pm0.12$ & $-1.305\pm0.015$\\  
$[8.0]$ &  $-1.462\pm0.099$ & $15.51\pm0.16$ & $-1.473\pm0.023$\\  
\hline
$W1$&                   &              & $-2.023\pm0.085$\\  
$W2$ &                  &              & $-1.516\pm0.056$\\  
 \hline
\end{tabular}
\tablenotetext{a}{The left-hand part of the table represents the
  best-fitting slopes and intercepts, while the right-hand part
  applies for DM = 15.506 mag.}
\end{center}
\end{table}

To verify the robustness of our method, we generated 46 artificial
Cepheids with DM = $15.506\pm0.108$ mag and a 1$\sigma$ spread as
observed. The Cepheids' color excesses, $E(\lambda-K_{\rm s})$,
covered the same ranges as the real data; $A_{\Ks}$ was estimated
based on the color excess multiplied by the slope (from Table
\ref{t2}). The uncertainties in the slopes were also taken into
account. We generated color excess vs. extinction diagrams based on
these artificial Cepheids (see Fig. \ref{f1}) to rederive the slopes
and intercepts. After repeating this process 10,000 times, we found
that the mean slope and intercept did not show any offset (to within
0.1\%) from the actual values. This confirms that our method is indeed
highly stable and suitable for determining Cepheid extinction values.

Based on the values for the slopes thus obtained, we can derive the
relative extinction, $A_{\lambda}/A_{\Ks}$: see Table \ref{t3}. To
ensure that the VVV and IRSF Cepheid samples are statistically
identical, we compared the relative extinction values resulting from
our analysis of the VVV and IRSF samples alone as well as those for
the combined sample. The result implies that they are in mutual
agreement given the prevailing systematic errors. In detail, the IRSF
sample is smaller and exhibits more significant scatter in the
distances compared with the VVV sample, so the result from the
combined sample is close to that based on the VVV sample only.

By fitting the NIR extinction values $A_J/A_{\Ks}$ and $A_H/A_{\Ks}$
for the combined sample, we derived the power-law index of the NIR
extinction law using the VVV effective wavelengths as our benchmarks,
$\alpha=2.05\pm0.05\pm0.05$. Here, the first error bar is the
uncertainty in the fit while the second uncertainty is the systematic
error. In addition, the IRSF and VVV photometric systems are
characterized by a small wavelength difference in the $H$ band. This
wavelength difference leads to 0.5\% difference in $\alpha$, which is
negligible compared with the statistical and systematic errors.

\begin{table}[h!]
\begin{center}
\caption{\label{t3}NIR and MIR relative extinction values.}
\vspace{0.15in}
\begin{tabular}{lccc}
\hline
\hline
      & VVV+IRSF & VVV & IRSF\\
\hline                                      
       $N$   &           46    &      34    & 23    \\ 
       \hline        
$A_J/A_{\Ks}$       & $3.005\pm0.031\pm0.094$ & $3.040\pm0.035\pm0.084$ & $2.880\pm0.065\pm0.166$\\   
$A_H/A_{\Ks}$       & $1.717\pm0.010\pm0.033$ & $1.734\pm0.010\pm0.030$ &$1.661\pm0.015\pm0.047$ \\   
$A_{[3.6]}/A_{\Ks}$ & $0.478\pm0.007\pm0.025$ & $0.480\pm0.008\pm0.026$&$0.491\pm0.009\pm0.035$\\   
$A_{[4.5]}/A_{\Ks}$ & $0.341\pm0.008\pm0.031$ & $0.338\pm0.008\pm0.027$&$0.381\pm0.010\pm0.043$\\   
$A_{[5.8]}/A_{\Ks}$ & $0.234\pm0.009\pm0.036$ & $0.233\pm0.010\pm0.031$&$0.264\pm0.011\pm0.050$\\   
$A_{[8.0]}/A_{\Ks}$ & $0.321\pm0.011\pm0.032$ & $0.326\pm0.013\pm0.028$&$0.340\pm0.013\pm0.058$\\   
$A_{W1}/A_{\Ks}$    & $0.506\pm0.022\pm0.026$ & $0.511\pm0.021\pm0.022$&$0.489\pm0.050\pm0.037$\\   
$A_{W2}/A_{\Ks}$    & $0.340\pm0.025\pm0.028$ & $0.343\pm0.036\pm0.027$&$0.344\pm0.040\pm0.050$\\
 \hline
\end{tabular}
\tablenotetext{a}{The three columns represent the relative extinction
  values determined based on the combined sample, the VVV sample, and
  the IRSF sample. The statistical and systematic errors are also
  indicated (in this order).}
\end{center}
\end{table}
\subsubsection{Uncertainties in the Relative Extinction}

To analyze the total uncertainties in the relative extinction, we
consider both the statistical and the systematic errors. Since we
combine seven bands to determine the optimal DM and extinction values,
the statistical errors are small (see Table 3). The systematic errors
are composed of two components, including the uncertainty in the DM
and the errors in the PLR zero points.\footnote{The effects of
  metallicity differences are not straightforward to estimate. 
    Based on the metallicity map of the bulge
    \citep{Gonzalez13}, the inner bulge's mean metallicity is similar to
    solar metallicity, while the dispersion is less than 0.08 dex in
    [Fe/H]. This adds a negligible uncertainty to the NIR and MIR
    PLRs.}
The uncertainty in the DM comes from simultaneously fitting the six
color excess--extinction diagrams in Section 3.1.1 and converting the
results to obtain the relative extinction. The DM uncertainties are
0.108, 0.086, and 0.169 mag for the combined sample, the VVV sample,
and the IRSF sample, respectively.
Since we use the PLRs to derive the absolute magnitudes and the
intrinsic colors, deviations of the PLR zero point will cause
absolute-magnitude and intrinsic color biases. The bias in the
absolute magnitude will affect the average DM, while the intrinsic
color bias will affect the color excess; both will affect the relative
extinction. The zero-point difference in the $K_{\rm s}$-band PLR is
evaluated on the basis of 10 Cepheids with {\sl Hubble Space
  Telescope} parallaxes \citep{Benedict07}. We determined a systematic
uncertainty of $\pm0.105/\sqrt{10}\approx 0.033$ mag. This is typical
of the uncertainties in NIR and MIR PLR zero points
\citep{Freedman12}. The small uncertainty of about 0.033 mag in the
intrinsic colors contributes $0.033/\langle A_{\Ks}
\rangle\approx0.013$ to the error in the relative extinction, which
accounts for 30\% of the systematic error. The combined statistical
and systematic errors in the relative extinction are tabulated in
Table \ref{t3}.

\subsection{The Color Excess Method}

\begin{figure} [h!]
\centering
\vspace{-0.0in}
\includegraphics[angle=0,width=160mm]{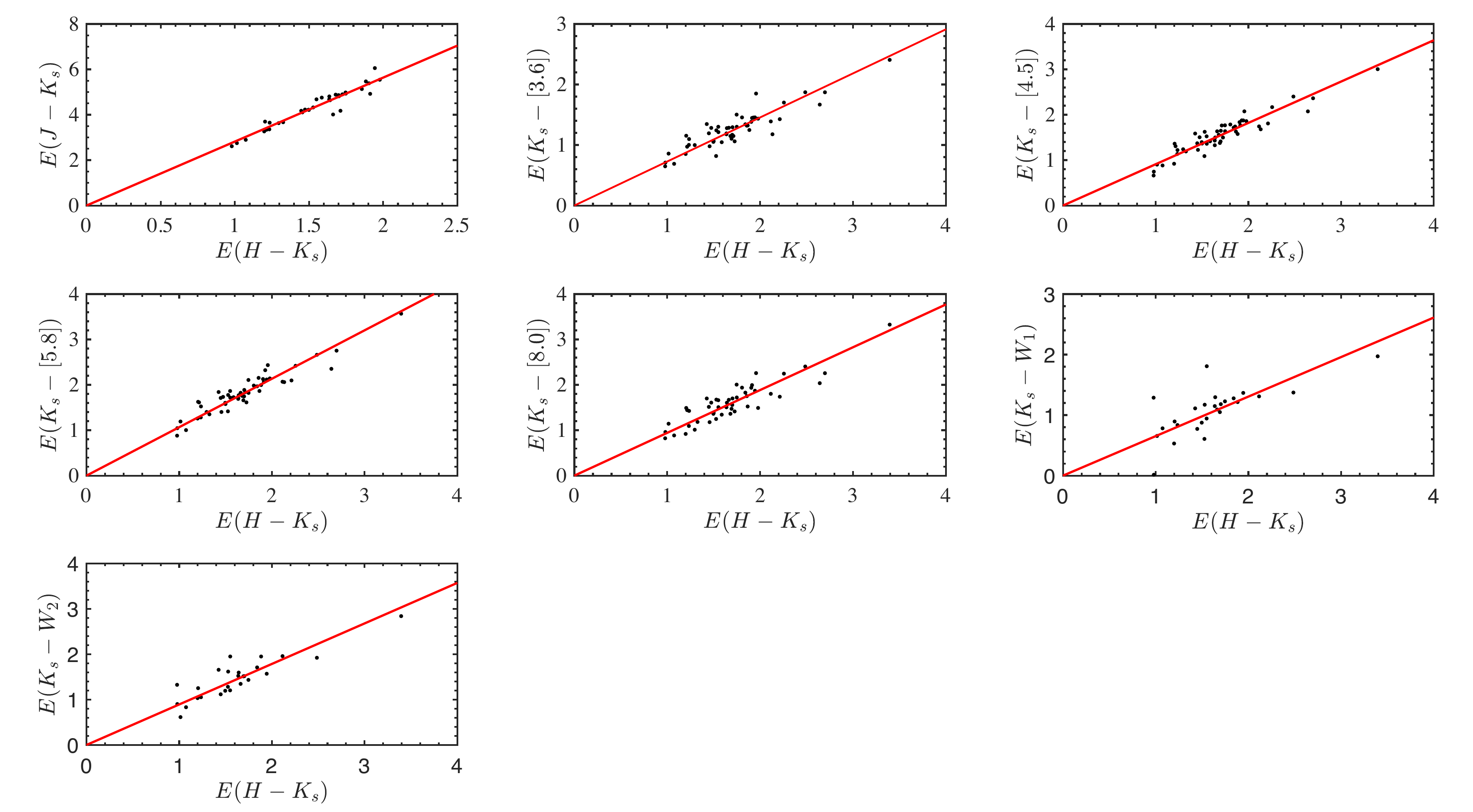}
\vspace{-0.1in}
\caption{\label{f5} Color excess--color excess diagrams for our 55
  Cepheids. The red lines are the best linear fits.}
\end{figure}

Color excess ratios (CERs), e.g., $E(\lambda-\Ks)/E(H-Ks)$, are also
usually used to express extinction variations. In extinction studies,
calculation of a CER is referred to as the color excess method. In a
color excess--color excess diagram composed of a group of tracers, the
CER is given by the slope of the linear fit. This method is suitable
for extinction determination to objects located at different
distances, which is an important difference with respect to the color
excess--extinction method discussed in Section 3.1. Therefore, we
computed the CERs for our 55 Cepheids to investigate the NIR--MIR
extinction law pertaining to the Galactic Center. The intrinsic colors
were calculated based on the PLRs, and the color excesses were
subsequently derived by comparison with the mean observed
magnitudes. Figure \ref{f5} shows the color excess--color excess
diagrams for different combinations of passbands. We determined
$E(J-\Ks)/E(H-\Ks)=2.82\pm0.02$, $E(\Ks-[3.6])/E(H-\Ks)=0.73\pm0.01$,
$E(\Ks-[4.5])/E(H-\Ks)=0.91\pm0.01$,
$E(\Ks-[5.8])/E(H-\Ks)=1.07\pm0.01$,
$E(\Ks-[8.0])/E(H-\Ks)=0.94\pm0.02$, $E(\Ks-W1)/E(H-\Ks)=0.65\pm0.03$,
and $E(\Ks-W2)/E(H-\Ks)=0.89\pm0.03$. By assuming either
$A_{H}/A_{\Ks}$ or a power-law index for the NIR extinction, the
relative extinction can be determined from the CERs.
Since we only intend to compare the extinction results with those
derived based on the color excess--extinction method, we adopt
$A_{H}/A_{\Ks}=1.717$, as determined in Section 3.1. The resulting
relative extinction values are $A_J/A_{\Ks}=3.02\pm0.03$,
$A_{[3.6]}/A_{\Ks}=0.48\pm0.01$, $A_{[4.5]}/A_{\Ks}=0.35\pm0.01$,
$A_{[5.8]}/A_{\Ks}=0.23\pm0.01$, $A_{[8.0]}/A_{\Ks}=0.32\pm0.02$,
$A_{W1}/A_{\Ks}=0.53\pm0.03$, and $A_{W2}/A_{\Ks}=0.36\pm0.02$.
Comparing these values with those listed in Table \ref{t3}, we find
that the relative extinction values are in good agreement with one
another. This means that the extinction law does not vary artificially
as a result of applying either one of these methods.

\subsection{Absolute Extinction in the Galactic Center}

In Sections 3.1 and 3.2, we estimated the NIR and MIR extinction laws
for Cepheids projected along sightlines toward the Galactic Center
region. For comparison, we also determined the extinction for the four
Cepheids located in the Galactic Center's nuclear disk ($-0.1^\circ
\lesssim l \lesssim +0.1^\circ$, $-0.1^\circ \lesssim b \lesssim
+0.1^\circ$). For the four central Cepheids (GCC-a, -b, -c, and
-d),\footnote{These four Galactic Center Cepheids are marked 1, 2, 3,
  and 4 in the second column of Table 6.} the NIR $JHK_{\rm s}$ mean
magnitudes were obtained from \citet{Matsunaga16}, while their
distribution around the Galactic Center is shown in Figure 7 of
\citet{Matsunaga15}. The MIR {\sl Spitzer} data were collected from
the GLIMPSE II catalog and from {\sl Spitzer} Program ID 12023 (PI:
R. Benjamin). We performed photometric analysis of all {\sl Spitzer}
image data for these Cepheids. Both point response function (PRF)
fitting and aperture photometry were applied to the corrected basic
calibrated data (CBCD) using the Mosaicking and Point-source
Extraction (MOPEX) package.  As regards the PRF fitting, PRF Maps for
the different bands, for both the cryogenic and warm missions, were
adopted. Correction of the PRF fluxes was also done. For our aperture
photometry, a small, 2-pixel aperture was used; aperture corrections
were obtained and adopted from the IRAC Data Handbook.

The fluxes resulting from both methods were comparable to within 2\%
precision, except for GCC-b. For this latter Cepheid, the aperture
flux is 20--50\% larger than the corresponding PRF flux. Based on a
check of the images, we found that this object's flux may be affected
by that of two nearby stars, located within $1''$ (or 0.8 pixels) from
the Cepheid. We excluded this Cepheid from our MIR analysis, since we
found (based on extensive experimentation) that the PRF fitting method
was unable to deliver a clean set of photometric measurements and
still overestimated its flux. The photometric magnitudes in the four
{\it Spitzer} bands for the remaining Cepheids---GCC-a, GCC-c, and
GCC-d---are listed in Table \ref{t5}, along with their period
phases. {\bf To evaluate the impact of background contributions on the
  stellar photometry, the background-to-star light flux ratios are
  listed after the magnitudes. Note that the background contributions
  have a negligible influence on the $[3.6], [4.5]$ fluxes, a small
  effect on the $[5.8]$ fluxes, and a more significant impact on the
  $[8.0]$ fluxes. As a result, the [8.0]-band magnitudes and the
  resulting extinction values will be associated with larger
  uncertainties and possibly a more significant scatter. If we would
  assume the background contributions to have the maximum impact on
  the photometry, their uncertainties propagate to relative extinction
  ratios $A_\lambda/A_{\Ks}$ of 0.004, 0.004, 0.020, and 0.107 in the
  [3.6], [4.5], [5.8], and [8.0] bands, respectively. These
  uncertainties are smaller than the total uncertainties (see below).}
The mean magnitudes were determined by adopting the NIR light curves
of \citet{Matsunaga15}.

Based on the apparent and absolute magnitudes for these objects, the
absolute extinction $A_{\lambda}$ ($J, H, \Ks$, [3.6], [4.5], [5.8],
and [8.0]) can be derived by adopting the recommended distance of $R_0
= 8.3\pm0.2\pm0.4$ kpc based on a statistical reanalysis of historical
publications \citep{deGrijs16}; see also
\citet{BlandHawthorn16}. Table \ref{t6} includes the NIR extinction
values for the four central Cepheids (based on IRSF measurements) and
the MIR extinction values for all Cepheids except for GCC-b. The error
in the mean relative extinction values contains both the statistical
and systematic uncertainties. We found that even if the variation in
the absolute extinction is large, e.g., $\sim$0.36 mag in the $\Ks$
band, the variation in relative extinction is small, $\sim$3\%. This
relative extinction value for the central Cepheids is comparable to
that for the other Cepheids projected close to the Galactic Center
(see Section 3.1). This means that the sightlines to the Galactic
Center and nearby regions obey similar NIR and MIR extinction laws. We
adopt NIR and MIR relative extinction values from Section 3.1.1 for
our subsequent distance analysis as these relative extinction errors
have been better evaluated (see Section 3.1.2).

\begin{table}[h!]
\begin{center}
\caption{\label{t5} {\it Spitzer} photometric results and mean Galactic
  Center Cepheid magnitudes}
\vspace{0.15in}
\begin{tabular}{lcccccc}
\hline
\hline
 Object &     BJD       &  Phase  &  [3.6]  &  [4.5]  &  [5.8]  &  [8.0]\\
        &    2400000+   &         &   mag   &   mag   &   mag   &   mag \\
GCC-a   &  57378.16358  &  0.02   &   9.04 (0.8\%)  &   8.74 (1.0\%)  &         &       \\
        &  57380.58993  &  0.12   &         &   8.51(0.8\%)  &         &       \\
        &  53459.31493  &  0.46   &   8.64  (0.6\%)&   8.38 (0.7\%)  &         &   8.04 (18.8\%) \\
        &  53460.60242  &  0.51   &   8.64  (0.6\%)&         &   8.04 (3.7\%)  &       \\
        &               &  Mean   &   8.84  &   8.54  &   8.24  &   8.24\\
\hline
GCC-c   &  53460.59990  &  0.08   &         &   8.51 (1.5\%) &         &   8.76 (35.5\%) \\
        &  53460.65728  &  0.08   &   8.92 (1.5\%)  &         &   8.39 (6.2\%) &       \\
        &  57379.10454  &  0.28   &         &   8.44 (1.4\%) &         &       \\
        &  57380.59232  &  0.35   &   8.74 (1.2\%)  &         &         &       \\
        &               &  Mean   &   8.88  &   8.53  &   8.35  &   8.73\\
\hline
GCC-d   &  57377.91906  &  0.21   &         &   8.64 (0.9\%) &         &       \\
        &  57379.10603  &  0.27   &   9.03 (0.8\%)  &   8.60 (0.8\%)   &         &       \\
        &  53459.11541  &  0.71   &   9.01 (0.8\%)  &         &   8.47  (4.1\%)&       \\     
        &  53460.65098  &  0.79   &         &   8.64  (0.8\%) &         &       \\
        &               &  Mean   &   9.06  &   8.65  &   8.52  &       \\
\hline
\end{tabular}
\end{center}
\end{table}

\begin{table}[h!]
\vspace{-0.0in}
\begin{center}
\caption{\label{t6}NIR and MIR extinction values for the Galactic
  Center based on the four nuclear Cepheids}
\vspace{0.15in}
\begin{tabular}{lccccc}
\hline
\hline
                           &    GCC-a & GCC-b & GCC-c & GCC-d &     mean       \\   
\hline                             
$A_{\Ks}             $    &    2.51  & 2.27  & 2.63  & 2.34  & $2.43\pm0.06$  \\     
$A_J/A_{\Ks}      $    &    2.99  & 3.11  & 3.11  &       & $3.07\pm0.08$  \\     
$A_H/A_{\Ks}      $    &    1.70  & 1.76  & 1.75  & 1.73  & $1.74\pm0.05$  \\     
$A_{[3.6]}/A_{\Ks}$    &    0.49  &       & 0.47  & 0.49  & $0.48\pm0.04$  \\     
$A_{[4.5]}/A_{\Ks}$    &    0.35  &       & 0.32  & 0.30  & $0.32\pm0.05$  \\     
$A_{[5.8]}/A_{\Ks}$    &    0.23  &       & 0.24  & 0.23  & $0.24\pm0.04$  \\     
$A_{[8.0]}/A_{\Ks}$    &    0.26  &       & 0.42  &       & $0.34\pm0.12$  \\     
\hline
\end{tabular}
\end{center}
\end{table}

\subsection{The Relation Between Color Excess Ratios and Galactic Longitude}

 \citet{Zasowski09} studied variations in the CER 
  $E(\lambda-\Ks)/E(H-\Ks)$ as a function of Galactic longitude,
  $l$, in the Galactic plane using RC stars. They covered
  sightlines in the ranges $10^\circ<l<65^\circ$,
  $265^\circ<l<350^\circ$, and $l=90^\circ$, thus avoiding the
  Galactic Center region. Our CERs based on classical Cepheids located
  at $-10.5^\circ \lesssim l \lesssim+10^\circ$ cover a region that
  is fully complementary to the longitude range considered by
  \citet{Zasowski09}.

Based on the estimated $E(\lambda-\Ks)/E(H-\Ks)$ ratios for individual
Cepheids, the mean values were determined in longitude bins of five degrees. 
The central value (i.e., for $l=0^\circ$) was determined 
based on the four nuclear Cepheids. 
For comparison, we convert our CERs to the 2MASS isophotal wavelength 
by application of factors of 1.077,1.033, 1.026, 1.023, and 1.026 
at $J$, [3.6], [4.5], [5.8], and [8.0].
These factors were estimated through ${\rm CER_{2MASS}/CER}$, 
assuming a NIR power law. 
As the influence of adopting different $\alpha$ is negligible, 
we take $\alpha=2.05$ in here. 
Although these corrections are crude, 
it allows us to better compare our results with previous publications.

\begin{figure} [h!]
\centering

\vspace{-0.5in}
\includegraphics[angle=0,width=140mm]{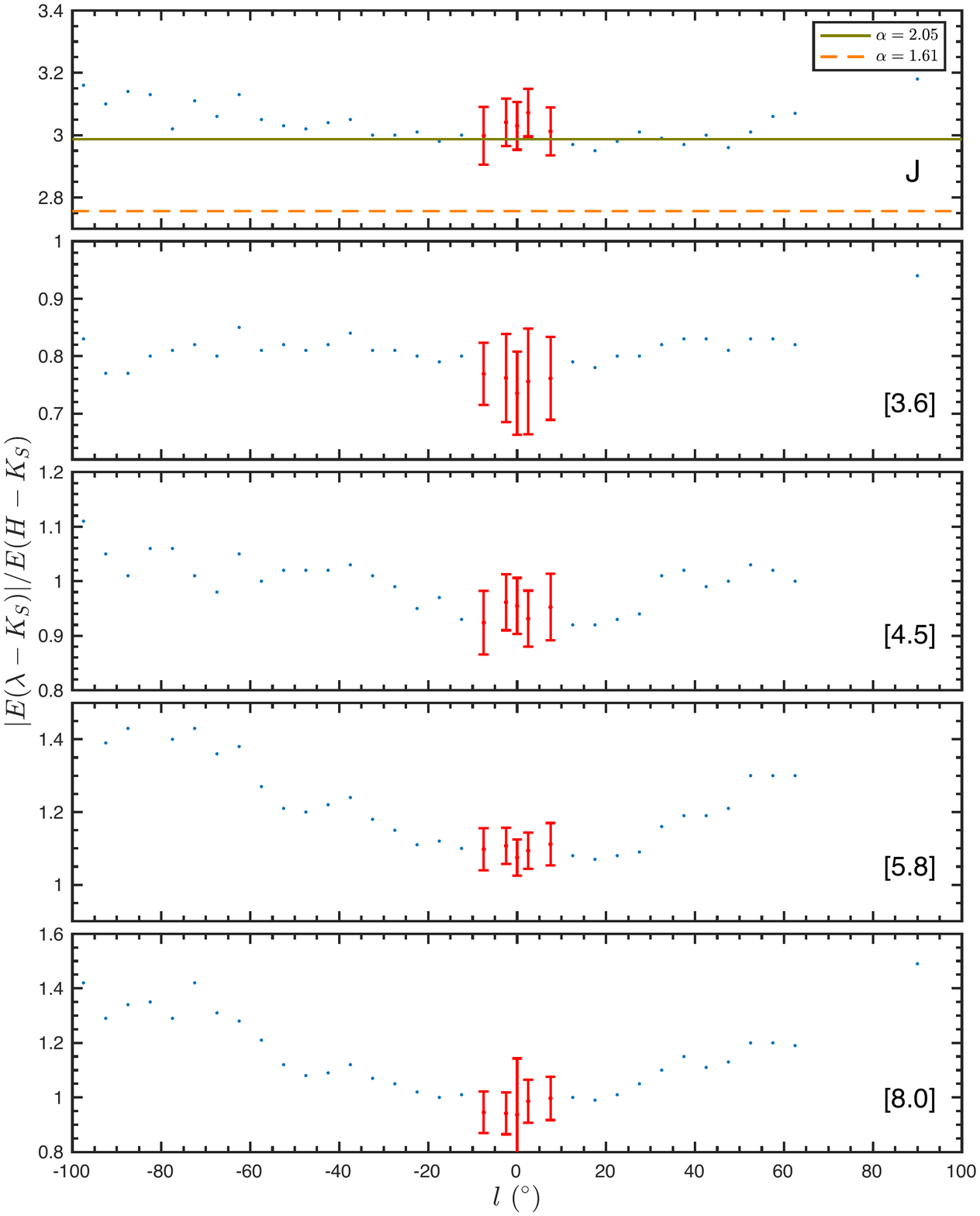}
\vspace{-0.5in}
\caption{\label{f6} Color excess ratios $E(\lambda-\Ks)/E(H-\Ks)$ vs. 
  Galactic longitude $l$ diagram in the $J$, [3.6], [4.5], [5.8], and [8.0] bands. 
  The blue dots are from \citet{Zasowski09}, while red dots are ours with
  $JH\Ks$ having been converted to 2MASS wavelengths. The light green
  and orange dashed lines are the $E(J-\Ks)/E(H-\Ks)$ ratios based on
  NIR power laws with $\alpha=2.05$ and $\alpha=1.91$, respectively.}
\end{figure}

Figure \ref{f6} shows CERs--$l$ diagram in the range of $-100^\circ<l<90^\circ$. 
Although the CERs vary as longitude globally, they do not vary significantly in the range 
$-20^\circ<l<20^\circ$ in either the NIR or the MIR regimes. 
As the relative extinction $A_\lambda/A_{\Ks}$ are also 
widely used to indicate extinction law, 
we also discuss the possible variations in $A_\lambda/A_{\Ks}$ by assuming $A_H/A_{\Ks}$. 
At sightlines of $-10^\circ \lesssim l \lesssim+10^\circ$, the variations in 
$A_{\lambda}/A_{\Ks}$ are 0.022, 0.008, 0.004, 0.014, 0.004, and 0.024 
for $\lambda=J, H, [3.6], [4.5], [5.8]$, and $[8.0]$, respectively. 
This is less than the statistical uncertainties in relative extinction 
listed in Table 3. At sightlines of $10^\circ < |l|<20^\circ$ \citep{Zasowski09}, 
the variations in $A_{\lambda}/A_{\Ks}$ are also small. 
Therefore, we conclude that the NIR and MIR extinction values 
do not vary appreciably in the Galactic Center region. 

Note, in the NIR, the variations in $E(J-\Ks)/E(H-\Ks)$ at 
$-100^\circ<l<90^\circ$ are much smaller than the differences 
resulting from assuming either $\alpha=1.61$ or $\alpha=2.05$. 
The variation is even smaller in the relative extinction $A_\lambda/A_{\Ks}$. 
For all of these CER values, by assuming $A_J/A_{\Ks}=3.12$, 
the corresponding $A_H/A_{\Ks}$ ratios are found to lie in 
the narrow range [1.667, 1.719], with $\sigma=0.014$. 
If we assume $A_H/A_{\Ks}=1.71$, the derived $A_J/A_{\Ks}$ ratios 
cover the narrow range [3.095, 3.258], with $\sigma=0.044$. 
Compared with the distance uncertainties (3--5\%) or 
the prevailing photometric uncertainties (1--2\%), 
the relative extinction ratios $A_H/A_{\Ks}$ (1\% uncertainty) 
do not vary across the range $-100^\circ<l<90^\circ$. 
This suggests that adoption of a typical $A_H/A_{\Ks}$ value 
determined in the Galactic Center region is more appropriate to 
determine the extinction law for inner Galactic plane (Section 5.2) than deriving 
the ratio based on a NIR power law with a poorly constrained value of $\alpha$.

We also examined possible variations in the CER along the color excess. 
The linear correlation coefficient relating
$E(J-\Ks)/E(H-\Ks)$ and $E(J-\Ks)$ is $R^2=0.004$. 
The CER thus does not vary with increasing extinction either.

\section{Cepheid Distances and Distribution}

Since the relative extinction in the infrared does not vary with
  Galactic longitude $l$ or as a function of increasing extinction, it
  can be used to determine the distance to any given Cepheid.
Previous Cepheid studies only used the $H$ and $\Ks$ bands to derive
the extinction, and they only relied on the $\Ks$-band to estimate the
resulting distances. This approach is prone to introducing large
errors in the resulting extinction values and distances, since both
parameters depend on the accuracy with which we know both the
empirical extinction law (worse than 5\% in the Galactic bulge) and
the PLRs (5\%). To reduce the errors, a multi-passband optimal
extinction and distance method is needed. Such a method is very
effective in improving the distance accuracy of not only Cepheids
\citep{Madore17, Wang18} but also of other distance tracers
\citep{Chen16}.

\subsection{Cepheid Distances: High Accuracy}

In this paper, we combined photometry in seven passbands ($J, H, \Ks$,
[3.6], [4.5], [5.8], and [8.0]) to refine our extinction estimates and
obtain the distances. The $W1$ and $W2$ bands were not used, since few
sources were detected in these filters, which additionally are
affected by significant photometric uncertainties. If we allow
$A_{\Ks}$ to vary, $A_{\lambda}$ ($\lambda$: $J, H$, [3.6], [4.5],
[5.8], and [8.0]) can be estimated based on the relative extinction
measurements. For each wavelength $\lambda$, the DM is derived as
DM$_{\lambda}=\langle m_{\lambda} \rangle - M_{\lambda} -A_{\lambda}$,
where $\langle m_{\lambda} \rangle $ is the mean observed
magnitude. The weighted mean DM is $\langle {\rm DM} \rangle= (\sum
{\rm DM}_{\lambda}/{{\sigma_\lambda}^2})/(\sum
{1/{\sigma_\lambda}^2})$, where $\sigma_\lambda$ is the mean magnitude
error in the $\lambda$ band. The weighted standard deviation
(statistical error) of the mean DM can then be derived, which depends
on the value of $A_{\Ks}$. The optimal DM and extinction $A_{\Ks}$ are
determined when the standard deviation is minimized. Our results are
collected in Table \ref{t4}, specifically in columns 3 and 4 (denoted
`C'). For comparison, the distances and extinction values of
\citet[][denoted `D']{Dekany15a, Dekany15b} and of \citet[][denoted
  `M']{Matsunaga13, Matsunaga16} are also included.

The uncertainty in our DM is composed of two components. The first of
those is the statistical error, which is represented by the maximum
value of either the standard deviation or the scatter in the PLR. We
discussed the standard deviation in the previous paragraph. The error
introduced by uncertainties in the PLR is $\sigma_{1}= \sqrt{(\sum{\rm
    {err}_{PLR}}^2/{{\sigma_\lambda}^2})/
  ((n-1)\sum{1/{\sigma_\lambda}^2})}$, where $\sigma_\lambda$ denotes
the error in the mean magnitude in each band and
$1/{\sigma_\lambda}^2$ is its weight (provided that they are
independent). If the errors are correlated, the relevant error is
represented by the smallest scatter in the seven PLRs. The real
uncertainties lie between these two values; we adopt the upper limit
to be conservative. The photometric uncertainty is calculated as
$\sigma_{2}=\sqrt{1/(\sum {1/{\sigma_\lambda}^2})}$; it is negligible.
The second component making up the total uncertainty in the DM is
composed of the systematic errors in the PLRs' zero points and the
extinction law. The bias associated with both the NIR and MIR PLR zero
points is 0.033 mag (see Section 3.1.2). The error associated with our
adoption of the extinction law is estimated in a similar fashion as
$\sigma_{1}$, but replacing $\rm err_{PLR}$ by $\rm err_{ext}$ and
changing $n$ from 7 to 6. 

The extinction errors, $\rm err_{ext}$, are taken to be the values
listed in Table \ref{t3}. Although the six-band extinction values
exhibit some correlations, which will increase $\rm err_{ext}$, the
NIR and MIR extinction values are anti-correlated, which ultimately
leads to a reduction in this uncertainty. We estimate the
  uncertainties by first assuming that our six extinction values are
  fully independent, then repeating the procedure on the assumption
  that they are fully interdependent. The resulting uncertainties are
  0.045 mag and 0.040 mag, respectively. To avoid underestimating our
  uncertainties, we adopt the larger extinction value.

In summary, as seen in column 3 of Table \ref{t4}, the first
uncertainty in our DM is around 0.07 mag, while the second is around
0.06 mag. For our sample of 55 classical Cepheids in the bulge, the
distance precision has thus been improved to 4\%. This represents a
significant improvement, since the distance uncertainties reported
previously are all greater than 7\%. A careful comparison with
\citet{Matsunaga13,Matsunaga16} shows that our DMs do not deviate
systematically from theirs (the average deviation is less than
  1\%); there is only 0.17 mag (8\%) statistical scatter between both
studies. However, compared with \citet{Dekany15a,Dekany15b}, we find a
27\% systematic difference in distance, which is due to the difference
in the adopted extinction law (see Section 4.1); the statistical
scatter is around 12\%.

\begin{center}
\begin{footnotesize}
\begin{longtable}{cccccccc}
\caption{DMs and $A_{\Ks}$ values for our 55 sample Cepheids based on
  the current study (denoted `C'), \citet[][denoted `D']{Dekany15a,
    Dekany15b}, and \citet[][denoted `M']{Matsunaga13,
    Matsunaga16}.}\label{t4}
\\
  \hline
R.A. (J2000) &  Dec. (J2000) &  DM(C) & $A_{\Ks}$(C) & DM(D) & $A_{\Ks}$(D) & DM(M) & $A_{\Ks}$(M)\\
(hh:mm:ss)    & $(^\circ  $ $'$  $''$ )       &  (mag)  & (mag)    & (mag)          & (mag)& (mag) & (mag)       \\
\hline
18:01:24.49 & $-$22:54:44.6 & $15.75\pm0.07\pm0.05$ & 2.79 & $15.27\pm0.18$ & 3.27 & 15.84 &    2.77\\
18:01:25.09 & $-$22:54:28.3 & $15.76\pm0.07\pm0.05$ & 2.71 & $15.28\pm0.18$ & 3.20 & 15.70 &    2.89\\
17:22:23.24 & $-$36:21:41.5 & $15.50\pm0.09\pm0.07$ & 1.72 & $15.20\pm0.19$ & 2.07 &        &     \\
17:21:16.05 & $-$36:43:25.2 & $15.31\pm0.07\pm0.06$ & 2.24 & $14.99\pm0.22$ & 2.59 &        &     \\
17:20:14.62 & $-$37:11:16.0 & $15.37\pm0.07\pm0.05$ & 1.42 & $14.90\pm0.21$ & 1.84 & 15.24 &    1.53\\
17:26:34.72 & $-$35:16:24.1 & $15.46\pm0.07\pm0.05$ & 2.34 & $14.90\pm0.17$ & 2.88 & 15.42 &    2.50\\
17:25:29.70 & $-$34:45:45.9 & $15.47\pm0.07\pm0.05$ & 2.33 & $15.05\pm0.17$ & 2.78 &        &     \\
17:26:43.41 & $-$34:58:25.6 & $15.91\pm0.07\pm0.06$ & 3.60 & $15.00\pm0.25$ & 4.50 &        &     \\
17:26:54.24 & $-$35:01:08.2 & $15.60\pm0.07\pm0.05$ & 2.79 & $15.17\pm0.19$ & 3.23 & 15.75 &    2.73\\
17:30:46.64 & $-$34:09:04.4 & $15.50\pm0.07\pm0.05$ & 2.53 & $14.88\pm0.19$ & 3.13 &        &     \\
17:28:15.86 & $-$34:32:27.2 & $15.21\pm0.07\pm0.06$ & 3.22 & $14.65\pm0.19$ & 3.78 &        &     \\
17:38:42.96 & $-$31:44:55.7 & $15.53\pm0.07\pm0.05$ & 2.36 & $15.12\pm0.17$ & 2.81 &        &     \\
17:36:44.46 & $-$32:04:38.6 & $15.28\pm0.09\pm0.06$ & 3.29 & $14.09\pm0.21$ & 4.40 &        &     \\
17:40:41.72 & $-$30:48:46.9 & $15.64\pm0.08\pm0.06$ & 2.34 & $14.98\pm0.20$ & 2.91 &        &     \\
17:40:25.15 & $-$31:04:50.5 & $14.89\pm0.07\pm0.05$ & 1.81 & $14.49\pm0.19$ & 2.19 &        &     \\
17:42:20.00 & $-$30:14:50.7 & $15.41\pm0.07\pm0.06$ & 3.51 & $14.79\pm0.22$ & 4.16 &        &     \\
17:41:15.13 & $-$30:07:17.7 & $15.41\pm0.09\pm0.05$ & 2.32 & $14.84\pm0.23$ & 2.84 &        &     \\
17:51:05.72 & $-$26:38:18.3 & $15.43\pm0.07\pm0.06$ & 2.45 & $14.92\pm0.18$ & 2.95 &        &     \\
17:51:13.77 & $-$26:48:55.9 & $15.55\pm0.07\pm0.05$ & 2.60 & $15.05\pm0.19$ & 3.11 & 15.74 &    2.45\\
17:49:41.42 & $-$27:27:14.6 & $15.41\pm0.07\pm0.06$ & 1.82 & $14.96\pm0.18$ & 2.25 & 15.40 &    1.87\\
17:50:30.49 & $-$27:13:46.7 & $15.54\pm0.07\pm0.05$ & 2.83 & $14.59\pm0.19$ & 3.70 & 15.20 &    3.06\\
17:53:16.07 & $-$26:28:26.9 & $15.90\pm0.24\pm0.05$ & 2.36 & $14.40\pm0.19$ & 3.62 &        &     \\
17:52:21.66 & $-$26:31:19.3 & $15.41\pm0.07\pm0.05$ & 2.73 & $14.94\pm0.19$ & 3.23 & 15.54 &    2.69\\
17:51:43.80 & $-$26:31:11.3 & $15.37\pm0.10\pm0.05$ & 2.74 & $14.45\pm0.19$ & 3.57 &        &     \\
17:55:44.68 & $-$25:00:30.2 & $15.51\pm0.07\pm0.06$ & 2.28 & $14.99\pm0.19$ & 2.77 &        &     \\
17:58:26.68 & $-$23:52:08.3 & $15.30\pm0.07\pm0.05$ & 2.63 & $14.80\pm0.21$ & 3.12 &        &     \\
18:03:31.12 & $-$22:21:14.0 & $15.51\pm0.07\pm0.05$ & 2.78 & $14.70\pm0.18$ & 3.55 &        &     \\
18:09:14.03 & $-$20:03:21.4 & $15.22\pm0.07\pm0.08$ & 4.72 & $14.31\pm0.21$ & 5.64 &        &     \\
17:22:10.10 & $-$36:44:18.8 & $15.38\pm0.07\pm0.06$ & 2.33 & $14.90\pm0.22$ & 2.77 &        &     \\
17:26:00.10 & $-$35:15:15.0 & $15.14\pm0.07\pm0.05$ & 2.44 & $14.72\pm0.18$ & 2.88 &        &     \\
17:32:14.07 & $-$33:23:59.5 & $15.46\pm0.07\pm0.04$ & 1.21 & $14.97\pm0.18$ & 1.69 & 15.22 &    1.44\\
17:40:51.51 & $-$30:24:53.2 & $15.35\pm0.07\pm0.05$ & 1.69 & $14.95\pm0.21$ & 2.11 &        &     \\
17:40:24.58 & $-$31:01:32.9 & $15.34\pm0.07\pm0.05$ & 2.19 & $15.12\pm0.19$ & 2.50 &        &     \\
17:50:17.54 & $-$27:08:13.3 & $15.61\pm0.07\pm0.06$ & 1.98 & $15.05\pm0.19$ & 2.47 &        &     \\
17:55:24.20 & $-$25:30:22.3 & $15.52\pm0.09\pm0.06$ & 2.29 & $15.05\pm0.19$ & 2.69 &        &     \\
17:54:40.25 & $-$25:34:39.5 & $15.34\pm0.07\pm0.06$ & 2.61 & $14.73\pm0.19$ & 3.17 & 15.33 &    2.61\\
17:56:01.96 & $-$25:15:44.9 & $15.43\pm0.07\pm0.06$ & 2.09 & $14.95\pm0.19$ & 2.55 &        &     \\
17:20:18.26 & $-$36:58:52.8 & $15.68\pm0.07\pm0.06$ & 1.61 &         &               & 15.53 &  1.70\\
17:24:12.58 & $-$36:01:46.9 & $15.50\pm0.07\pm0.05$ & 1.42 &         &               & 15.53 &  1.44\\
17:29:59.17 & $-$34:09:55.1 & $15.20\pm0.07\pm0.05$ & 1.32 &         &              & 15.14 &   1.40\\
17:38:46.14 & $-$31:26:22.8 & $15.43\pm0.08\pm0.05$ & 2.23 &         &              & 15.74 &   2.03\\
17:40:41.03 & $-$30:41:38.6 & $15.33\pm0.07\pm0.06$ & 2.09 &         &               & 15.28 &  2.14\\
17:44:56.91 & $-$29:13:33.8\footnotemark[4] & $14.56\pm0.07\pm0.05$ & 2.38 &         &               & 14.48 &  2.45\\
17:50:11.26 & $-$27:19:42.9 & $15.50\pm0.07\pm0.04$ & 1.96 &         &              & 15.19 &   2.17\\
17:52:28.94 & $-$26:23:40.0 & $15.62\pm0.07\pm0.07$ & 2.11 &         &               & 15.71 &  2.07\\
17:52:38.14 & $-$26:19:43.3 & $15.52\pm0.07\pm0.06$ & 1.86 &          &              & 15.71 &  1.77\\
17:57:31.41 & $-$24:30:26.7 & $15.57\pm0.10\pm0.06$ & 3.00 &         &               & 15.91 &  2.80\\
18:02:14.84 & $-$22:27:10.7 & $15.81\pm0.07\pm0.05$ & 2.61 &         &               & 15.88 &  2.58\\
18:03:29.93 & $-$22:03:22.5 & $15.97\pm0.07\pm0.06$ & 1.91 &         &               & 16.30 &  1.72\\
18:03:53.95 & $-$21:58:11.7 & $15.86\pm0.07\pm0.05$ & 2.20 &         &               & 15.87 &  2.22\\
18:05:52.84 & $-$21:06:41.9 & $15.86\pm0.07\pm0.05$ & 2.30 &         &               & 16.01 &  2.21\\
18:07:07.82 & $-$20:34:50.1 & $15.35\pm0.07\pm0.05$ & 2.26 &         &               & 15.14 &  2.42\\
17:46:06.01 & $-$28:46:55.1\footnotemark[1] & $14.62\pm0.07\pm0.08$ & 2.48 &         &               & 14.53 &  2.53\\
17:45:32.27 & $-$29:02:55.2\footnotemark[2] & $14.47\pm0.07\pm0.09$ & 2.39 &         &               & 14.28 &  2.50\\
17:45:30.89 & $-$29:03:10.5\footnotemark[3] & $14.52\pm0.07\pm0.09$ & 2.74 &          &              & 14.34 &  2.83\\

\hline
\end{longtable}
\end{footnotesize}
\end{center} 

\subsection{Cepheid Distribution}

\begin{figure}[h!]
\centering
\vspace{-0.0in}
\includegraphics[angle=0,width=120mm]{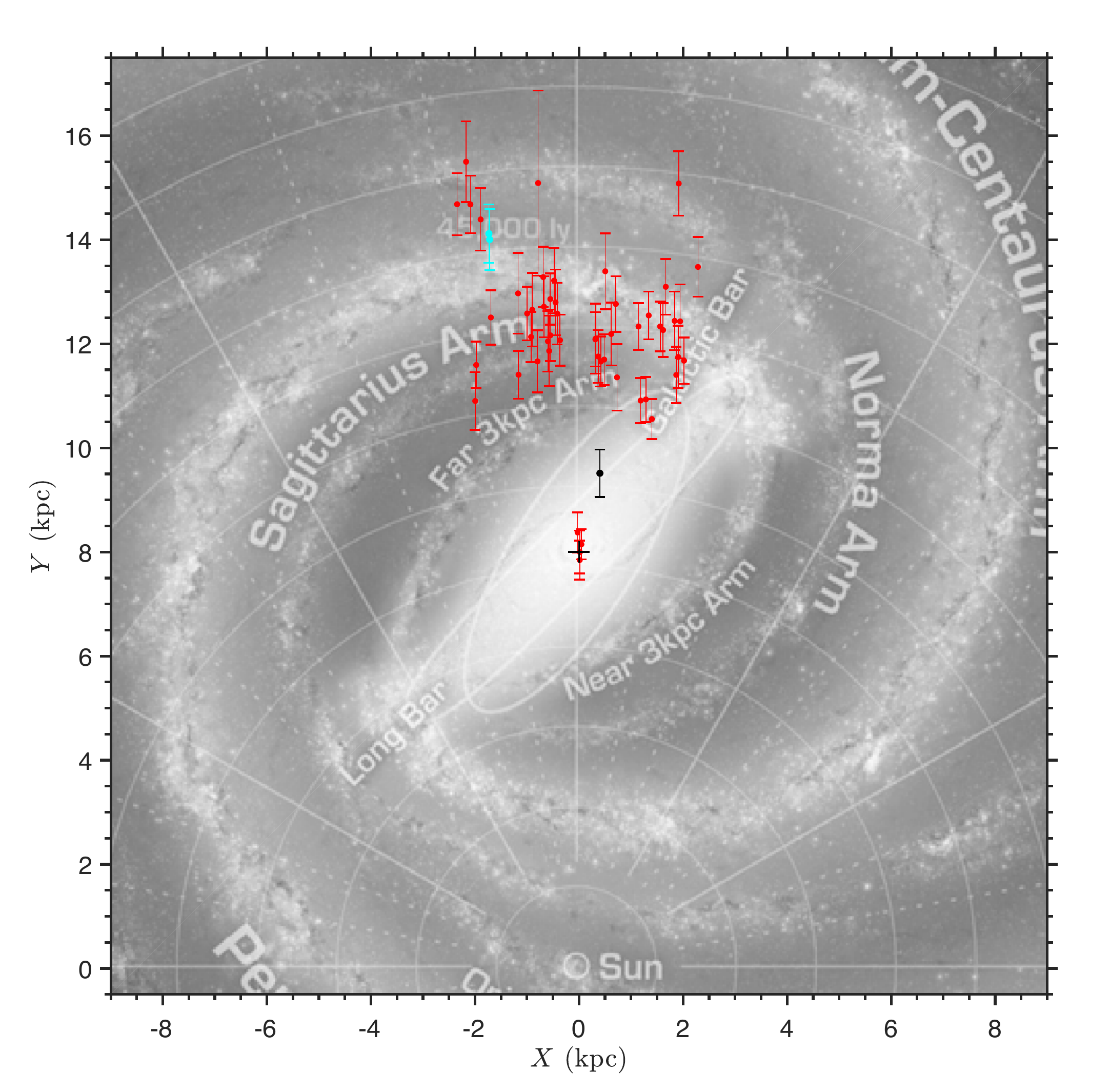}
\vspace{-0.0in}
\caption{\label{f4} Mapping of the 55 Cepheids superimposed on Robert
  Hurt's sketch of our Galaxy; $R_0=8.0$ kpc was adopted (see the
  black `+' sign). The cyan dots denote the position of
  \citet{Dekany15a}'s pair of Galactic Center Cepheids.}
\end{figure}

Figure \ref{f4} shows the spatial distribution of our 55 sample
Cepheids. Most are located at distances of about 12.5 kpc, which is
more distant than the average value of 9.5 kpc obtained by
\citet{Dekany15b}. Our sample Cepheids are apparently located in the
main and second spiral arms on the opposite side of the Galactic
Center. In the inner bulge, only four Cepheids are distributed across
the nuclear stellar disk. No other Cepheids tracing any other stellar
disk component are found in the inner bulge. Our results thus support
those of \citet{{Matsunaga16}} based on a direct method which properly
deals with the Galactic Center's heavy extinction.

 We point out the presence of a single Cepheid near the nuclear
  disk, which is unusual (see the black dot in Fig. \ref{f4}). It is
  located at a distance of $R=1.5\pm 0.4$ kpc from the Galactic Center
  (corresponding to a heliocentric distance of $9.5\pm0.4$ kpc). This
  Cepheid is located at $l=-2.4^\circ$, it has a period of 4.7 days,
  and its extinction is $\Ks=1.89\pm0.09$ mag. Given the heavy
  extinction, this Cepheid is unlikely a foreground Type II Cepheid
  (it would be located at a distance of 3.9 kpc if this were the
  case). It is likely that the object is a misidentified
  first-overtone Cepheid, for the following reasons. Its period of 4.7
  days falls within the overlap region of fundamental-mode (1.14 to
  52.9 days) and first-overtone Cepheids (0.27 to 5.91 days),
  according to \citet{Macri15}. Its distance modulus of $15.46\pm0.09$
  mag coincides with those of the other 46 Cepheids in the arm if we
  adopt the PLRs for first-overtone Cepheids. A $J$-band light curve
  or kinematic information are needed to achieve better constraints.
\subsection{The Galactic Center Distance}

As discussed in Section 3.3, four classical Cepheids are located in
the Galactic Center itself. Their distances were determined in Section
4.1 and are listed in Table 6. We take the mean distance of these four
Cepheids as the Galactic Center distance, DM = $14.541\pm0.050 \mbox{
  (\rm statistical)} \pm0.056 \mbox{ (\rm systematic)}$ mag
($R_0=8.10\pm0.19\pm0.21$ kpc). This distance value is reliable, since
it is based on seven-passband NIR and MIR data, and all potential
systematic errors have been taken into account.

Compared with the Galactic Center's fiducial distances derived based
on three of these four Cepheids by \citet{Matsunaga11},
$R_0=7.9\pm0.2\pm0.3$ kpc, and of all four Cepheids by
\citet{Matsunaga16}, $R_0=7.6\pm0.4 (\rm statistical)$ kpc, we have
derived a significantly reduced uncertainty.
Our Galactic Center distance based on Cepheids is more consistent with
the recommended distance of $R_0 = 8.3\pm0.2\pm0.4$ kpc based on a
statistical reanalysis of historical publications \citep{deGrijs16}.
However, using Cepheids to determine the distance has the advantage
that the systematic error is close to the true value, since the
systematic error is often underestimated in the literature. In the
previous sections, we have highlighted that the systematic error
pertaining to our Cepheid distance is a combination of contributions
by uncertainties in the extinction law (5\%) and the zero point of the
PLR (1.7\%). Even with our seven-band constraints, the total
  error cannot be smaller than 2.8\%. The actual error,
  $R_0=8.10\pm0.19\pm0.21$ kpc, is indeed very close to this limit.

\section{Discussion}
\subsection{The NIR Extinction Law: Underestimated Extinction Errors}

NIR relative extinction estimates are widely used to correct the
properties of the Galactic Center's Cepheids. However, adoption of
different extinction values will lead to large differences in the
Cepheids' distances. \citet{Dekany15a, Dekany15b} adopted
$A_{\Ks}/E(H-\Ks)=1.63\pm0.04$ \citep{Nishiyama09} to correct for the
$K$-band extinction, which leads to a systematic difference in
distance of at least 15\% with respect to the results of
\citet{Matsunaga13, Matsunaga16}, who used
$A_{\Ks}/E(H-\Ks)=1.44\pm0.01$ \citep{Nishiyama06}. The former value
of 1.63 was derived from the relative extinction
$A_H/A_{\Ks}=1.62\pm0.04$ \citep{Nishiyama09}, while the latter value,
1.44, was based on $A_H/A_{\Ks}=1.73\pm0.03$ \citep{Nishiyama06}. In
fact, the uncertainty in the relative extinction only includes the
statistical component. If we only assume a small systematic extinction
error of $\sigma \sim 0.05$ mag, the total uncertainty propagating to
$A_{\Ks}/E(H-\Ks)$ is $\pm0.18$ for a central value of 1.63. This
error almost fully encompasses the 15\% distance
difference.\footnote{$A_H/A_{\Ks}$ differs by only 1\% between
  different systems, e.g., between VVV and SIRIUS. This difference is
  much smaller than either the systematic or the statistical errors.}
In addition, \citet{Nishiyama09} claim that both results are similar
within about 2$\sigma$. This suggests that a systematic error must be
present that is somewhat larger than their statistical error of 0.04.
We derived the NIR extinction law for the Galactic Center based on
three different approaches. Our best estimate is
$A_{\Ks}/E(H-\Ks)=1.39\pm0.07$, which is close and within the
1$\sigma$ uncertainty to the central value of 1.44 used by
\citet{Matsunaga13, Matsunaga16}.

The distance precision of classical Cepheids is limited by the
prevailing uncertainties in the extinction law, the PLR, the
metallicity, and the photometric errors. For both Galactic and
extragalactic Cepheids, the NIR and MIR extinction errors are usually
neglected, while uncertainties in the metallicity are prominent
\citep{Freedman12}. However, for Cepheids observed toward the Galactic
Center, the extinction error $\sigma_{\rm {ext}}$ represents the most
prominent contribution, affecting the resulting distances by at least
5\%.
To ensure that the distance to Galactic bulge objects is unbiased, a
careful analysis of the systematic errors affecting the extinction law
is urgently needed. However, not only the color excess method
pertaining to RGs and RC stars, but also the color excess--extinction
method of relevance to specific RC stars is based on a number of
important underlying assumptions. This thus renders a systematic error
analysis complicated. Based on our analysis of Cepheids with
well-studied properties, we have determined the NIR--MIR extinction
law toward the Galactic Center, for the first time also including a
complete error analysis (see Section 3.1.2).

\begin{figure}[h!]
\centering
\vspace{-0.3in}
\includegraphics[angle=0,width=115mm]{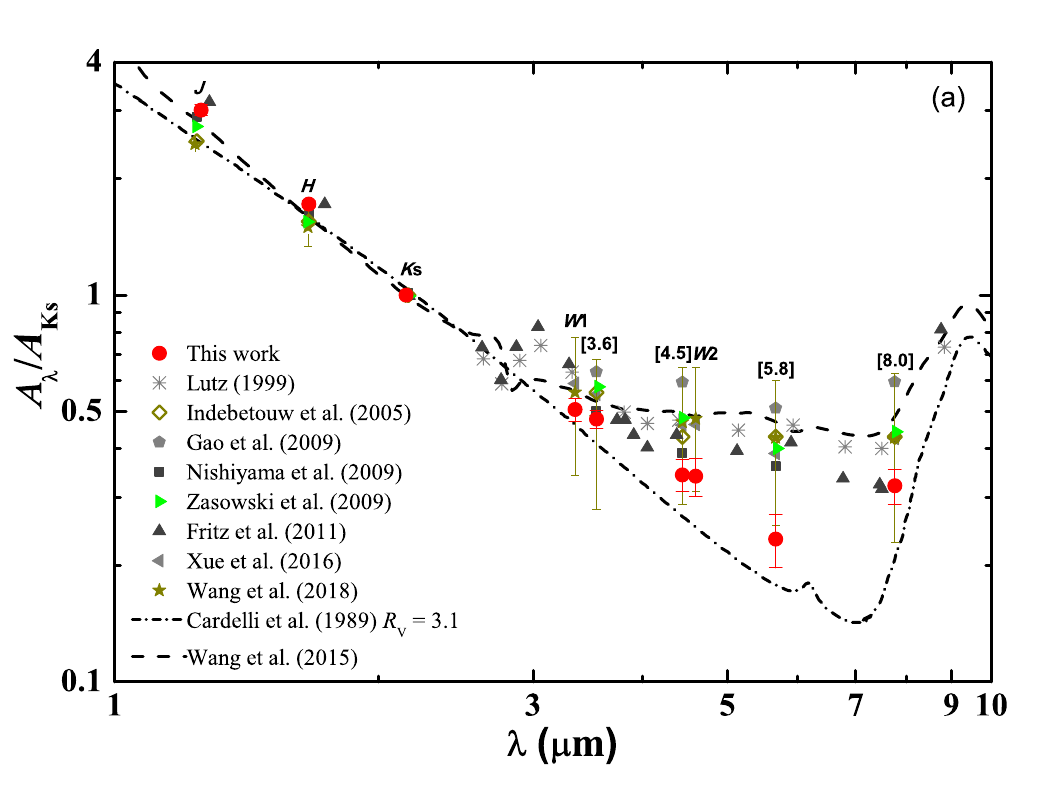}
\vspace{-0.2in}
\vspace{-0.2in}
\includegraphics[angle=0,width=115mm]{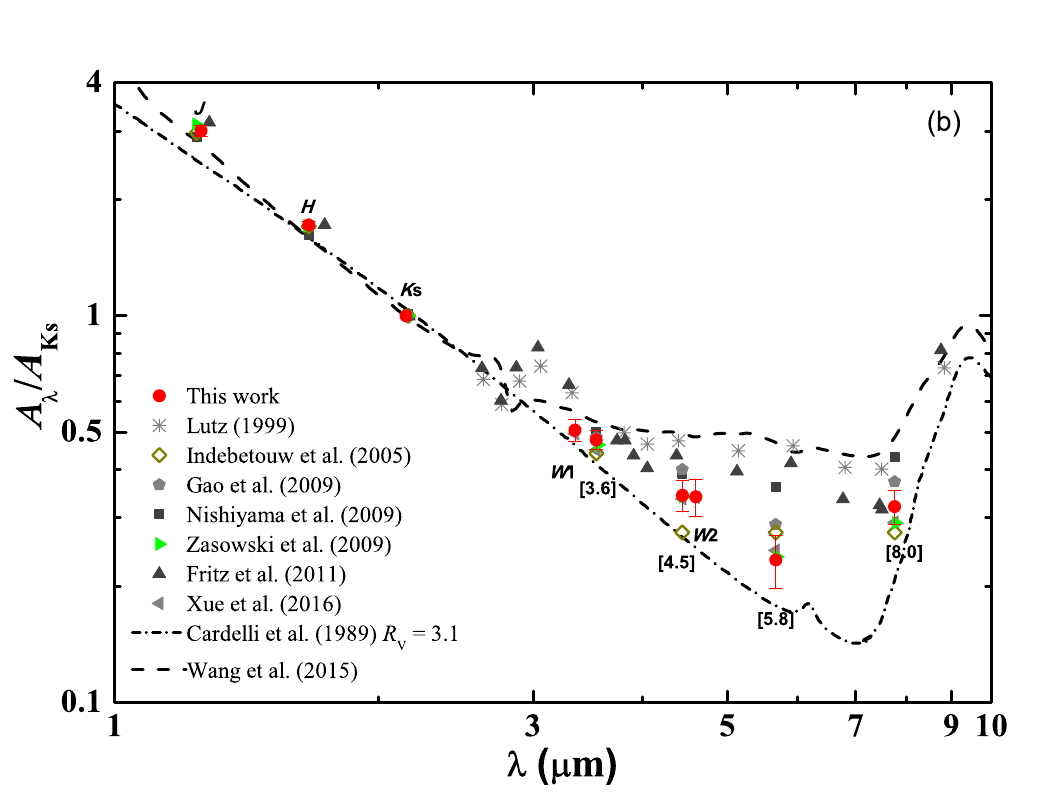}
\vspace{-0.1in}
\caption{\label{f2} Relative extinction diagram. The dash-dotted line
  represents $\Rv=3.1$ for a steep MIR extinction law
  \citep{Cardelli89}, while the dashed line is a flat MIR extinction
  law produced by the model of \citet{Wang15}. The red dots represent
  the eight-band relative extinction derived in this paper. We compare
  this law with a number of previous results (represented by different
  symbols), which tend to follow a flat MIR extinction law. Extinction
  of the four studies \citep{Indebetouw05, Zasowski09, Gao09, Xue16}
  shown in panel (b) are uniformly determined based on
  $A_H/A_{\Ks}=1.703$, while panel (a) show their values. }
\end{figure}
\subsection{The MIR Extinction Law: Extremely Low in the Galactic Center}

Compared with the extinction in the NIR, the extinction behavior at
MIR wavelengths of $\sim$3--8 $\mu$m is more complicated. In the
twentieth century, the MIR extinction law was thought to be an
extension of the NIR power law, $A_\lambda\propto \lambda^{-1.61}$,
out to $\sim$7 $\mu$m, i.e., before the extinction becomes dominated
by the 9.7 $\mu$m silicate absorption feature \citep{Rieke85}. This
was subsequently confirmed by \citet{Bertoldi99} and
\citet{Rosenthal00}, who derived $\alpha=1.7$ based on studies of the
ro-vibrational emission lines of H$_2$ in the Orion Molecular
Cloud. In addition, \citet{Draine89} pointed out that the
silicate--graphite model for diffuse clouds with $\Rv=3.1$ predicts
that the mid-IR extinction is a continuation of the near-IR power law
with $\alpha=1.7$ \citep{Weingartner01}.
However, \citet{Lutz96} derived the extinction from 2.5 $\mum$ to 9
$\mum$ toward Sgr A$^{\ast}$ in the Galactic Center based on atomic H
recombination lines. They found that the Galactic Center extinction in
the $\sim$3--8 $\mu$m region exhibits a flattened extinction law,
distinct from the projected continuation of the NIR power law and
lacking the pronounced minimum expected based on the
silicate--graphite model. This was later confirmed by \citet{Lutz99},
\citet{Nishiyama09}, and \citet{Fritz11}. Moreover, in the current
century an increasing number of studies suggest that the MIR
extinction law in both diffuse and dense environments departs
significantly from the NIR power law \citep{Indebetouw05, Flaherty07,
  Gao09, Wang13, Xue16}.

As shown in Fig. \ref{f2}(a), compared with previously published
results for different sightlines, our MIR relative extinction values
(red dots) are the lowest data points, in particular in the context of
the extremely low extinction in the [5.8] band,
$A_{[5.8]}/\AKs=0.234$. That latter data point is located far from the
flat MIR extinction curves represented by the different symbols in
Fig. \ref{f2}(a), e.g., $A_{[5.8]}/\AKs \approx 0.427$ of Wang et
al.\ (2018) for the Galactic plane based on classical Cepheids and
$A_{[5.8]}/\AKs \approx 0.389$ of Xue et al.\ (2016) for the whole
sky, based on G- and K-type giants, and close to the average
extinction law representative of sightlines toward diffuse Galactic
clouds for $\Rv =3.1$ ($A_{[5.8]}/\AKs \approx 0.19$).

 The main reason why our low MIR extinction values deviate from
  other extinction values obtained in the Galactic plane
  \citep{Indebetouw05, Gao09, Zasowski09, Xue16} resides in the
  $A_{H(J)}/A_{\Ks}$ ratios assumed. The CER method requires a
  fiducial relative extinction ratio $A_{H(J)}/A_{\Ks}$. The values
  adopted for this ratio in the four studies used for comparison are
  $A_J/A_{\Ks}=2.50$, $A_J/A_{\Ks}=2.52$, $A_H/A_{\Ks}=1.55$, and
  $A_J/A_{\Ks}=2.72$. If these values are adjusted to the absolute NIR
  relative extinction in the Galactic Center, $A_J/A_{\Ks}=3.119$ and
  $A_H/A_{\Ks}=1.703$ in 2MASS system 
  (converted from our values $A_J/A_{\Ks}=3.005$
  and $A_H/A_{\Ks}=1.717$ in VVV system), 
  the same low and steep MIR extinction law
  we find is also found in the four comparison studies (see Fig. \ref{f2}b). 
  The two studies that are based on RC stars and
  RGs with careful intrinsic color correction 
  \citep{Zasowski09, Xue16} are fully in accordance with our results.  
  Considering the uncertainties, the corrected values of 
  \citet{Indebetouw05} are also comparable with our results. 
  \citet{Gao09} derived a value that is somewhat higher than ours, 
  but this may also be driven by the lack of an intrinsic color correction.

Combined with our conclusion in Section 3.4, we find that the relative
extinction $A_H/A_{\Ks}$ is approximately uniform across all
sightlines in the inner Galactic plane. Adoption of $A_H/A_{\Ks}$ derived in
the Galactic Center region is more appropriate than using the ratio
determined based on adoption of power law index $\alpha \sim1.6$. 
This will soon be validated by {\sl Gaia}, since this latter mission 
will provide distance constraints to numerous RC stars. 
The extinction values of \citet{Wang18} were determined based on nearby Cepheids with $A_{\Ks}$ = 0.10--0.63 mag. 
The low extinction found here will be
affected by the large uncertainties in distances and photometric
measurements in most other studies, which consequently may hide the
MIR trend we derived.
 
In comparison with previous extinction results toward the Galactic
Center, our values are even lower than those of \citet{Lutz99},
\citet{Nishiyama09}, and \citet{Fritz11}, which are in turn less flat
than those pertaining to other regions.  The extinction estimate of
\citet{Nishiyama09} was based on RC stars in the Galactic bulge. Their
selection of RC stars considered the relevant density profile; it is
inevitably contaminated by foreground RC stars.  Extinction owing to
foreground RC stars would lead to the integrated relative extinction
in the MIR flattening. On the other hand, our extinction estimate in
this paper is based on 55 classical Cepheids with confirmed distances.
Therefore, it represents a more realistic MIR extinction estimate in
the Galactic bulge. We will discuss the extinction along the
  Galactic Center sightline in the next section.

\subsection{The IR Absolute Extinction in the Galactic Center}

\begin{figure}[h!]
\centering
\vspace{-0.0in}
\includegraphics[angle=0,width=140mm]{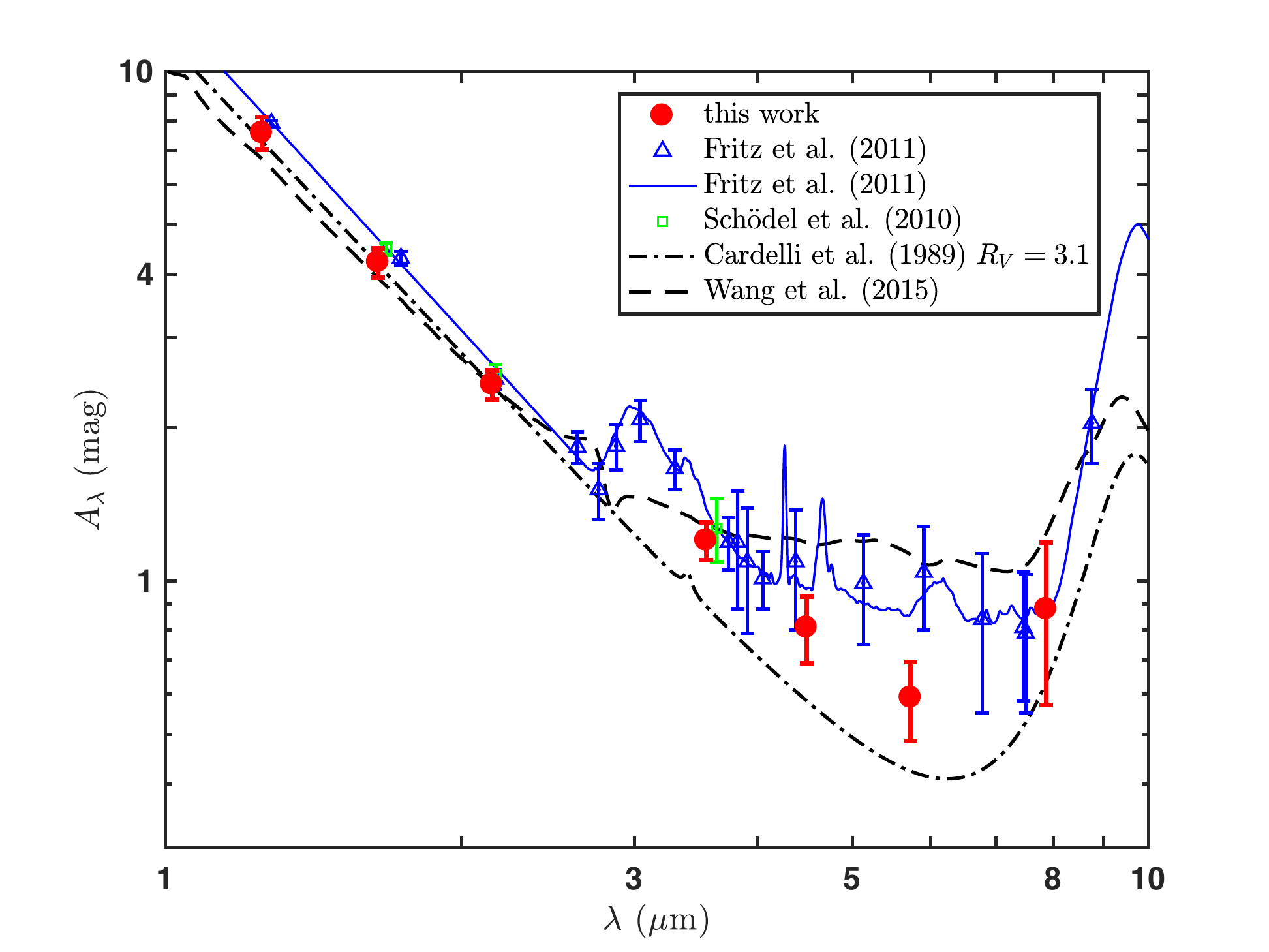}
\vspace{0.0in}
\caption{\label{f3} Absolute NIR--MIR extinction law in the Galactic
  Center. The red dots represent the seven-band extinction for the
  four confirmed central Cepheids. The blue triangles and blue line
  denote the H{\sc i} extinction and extinction curve from
  \citet{Fritz11}. The green squares are extinction values based on
  RC stars from \citet{Schodel10}. The dash-dotted and dashed
  lines are based on the relative extinction laws of
  \citet{Cardelli89} for $\Rv=3.1$ and \citet{Wang15}, respectively,
  for $A_{\Ks}=2.42$ mag.}
\end{figure}

Since an assessment of the absolute extinction is meaningful, we use
the four confirmed Galactic Center Cepheids as our basis for a
discussion. With the mean $\Ks$-band absolute extinction derived in
Section 3.3 (Table 5), the absolute extinction values for the Galactic
Center can be determined, i.e., $A_{J}=7.57\pm0.56, A_{H}=4.22\pm0.28,
A_{\Ks}=2.43\pm0.16, A_{[3.6]}=1.20\pm0.10, A_{[4.5]}=0.81\pm0.12,
A_{[5.8]}=0.59\pm0.10$, and $A_{[8.0]}=0.88\pm0.31$ mag. These
  values should be compared with $A_{H}=4.48\pm0.13$ mag,
  $A_{\Ks}=2.54\pm0.12$ mag, and $A_{L'}=1.27\pm0.18$ mag (the central
  wavelength of the $L'$ band is 3.64 $\mu$m) from \citet{Schodel10},
  who determined the extinction values based on RC stars. The
  two sets of results are in mutual accordance considering the
  uncertainties. If we correct their adopted Galactoc Center distance
  to our value of $R_0=8.3$ kpc (corresponding to a 0.07 mag
  difference in the distance modulus), $A_{L'}=1.20\pm0.18$ mag is the
  same as our extinction measurement in the [3.6] band. This means
  that the absolute extinction values in the Galactic Center based on
  photometric methods are similar. We also compare our results with
  spectroscopic methods based on the H{\sc i} line \citep{Fritz11}.
In the NIR, this is consistent with the determinations of
$A_{H}=4.21\pm0.10$ mag and $A_{\Ks}=2.42\pm0.10$ mag by
\citet{Fritz11}. In the MIR, and considering the uncertainties in the
extinction estimates of \citet{Fritz11} ($\sim$25\%, except for the
[5.8] band), our results are comparable with theirs: see Figure
\ref{f3}. For comparison, we also converted the relative extinction
laws of \citep{Cardelli89} for $\Rv=3.1$ and \citet{Wang15} to
absolute extinction values by adopting $A_{\Ks}=2.42$ mag (see,
respectively, the dash-dotted and dashed lines in Fig. \ref{f2}).
In addition, our absolute extinction law is close to the steep
$\Rv=3.1$ MIR extinction law. We are confident that the steep MIR
extinction is realistic and reliable. This is so, because if we adopt
a flat MIR extinction law with $A_{[5.8]}/A_{\Ks}=0.4$ or
$A_{[5.8]}=1.0$, the four central Cepheids would reside at a distance
of $R_0 = 6.5$ kpc. This distance differs by 20\% with respect to the
current best value.

\section{Conclusions}

We have collected 55 Cepheids from \citet{Matsunaga11, Matsunaga16}
and \citet{Dekany15a, Dekany15b} to study the extinction toward and
the distance to the Galactic Center. Three different approaches were
adopted to estimate the NIR--MIR extinction law. The good mutual
consistency among the three methods means that the extinction varies
little in and around the Galactic Center. Systematic errors affecting
the extinction laws were also discussed in detail. The resulting
relative extinction values are $A_J/A_{\Ks}=3.005\pm0.031\pm0.094,
A_H/A_{\Ks}=1.717\pm0.010\pm0.033,
A_{[3.6]}/A_{\Ks}=0.478\pm0.007\pm0.025,
A_{[4.5]}/A_{\Ks}=0.341\pm0.008\pm0.031,
A_{[5.8]}/A_{\Ks}=0.234\pm0.009\pm0.036,
A_{[8.0]}/A_{\Ks}=0.321\pm0.011\pm0.032,
A_{W1}/A_{\Ks}=0.506\pm0.022\pm0.026$, and
$A_{W2}/A_{\Ks}=0.340\pm0.025\pm0.028$ mag.

The MIR extinction law we derived here is lower than most of those
proposed in previous studies to the Galactic Center. It is close
to the steep $R_V=3.1$ MIR extinction law of \citet{Cardelli89}. The
absolute MIR extinction in the Galactic Center is lower than that
found by \citet{Fritz11}, especially in the [5.8] band. We also
  suggest that a relative extinction $A_H/A_{\Ks}=1.703$ may be better
  to determine the extinction in the inner Galactic plane. In turn, this
  also returns a steep MIR extinction law. Based on the newly derived
extinction law and using known Cepheid PLRs, we used a seven-band
optimal distance method to improve the distances to our sample of 55
Cepheids, resulting in a 4\% overall precision. Our distance
distribution of Cepheids supports the suggestion of
\citet{Matsunaga16}: except for the nuclear disk, no other disk
appears to exist in the inner bulge.

The systematic error in the distance is at least 5\%, based on
uncertainties in the extinction law, plus 1.7\% from zero-point errors
in the PLRs. Based on our multi-band constraints, the lower limit to
the total error is 2.8\%. A distance of $R_0=8.10\pm0.19\pm0.21$ kpc
was determined based on the four confirmed Galactic Center
Cepheids. This distance is in accordance with the statistical value
based on hundreds of previous publications, $R_0 = 8.3\pm0.2\pm0.4$
kpc \citet{deGrijs16}. With the prospects of new {\sl Gaia} data,
which will cover 9000 Cepheids, the scatter in the MIR PLRs will be
reduced to 0.06 mag or less, while the zero-point uncertainty will be
less than 1\%. The extinction law will benefit even more, since the
accuracy of many methods to derive extinction is limited by the lack
of independent distances. Consequently, the Galactic Center distance
precision will improve to 1--2\%.

\acknowledgments{We sincerely thank the anonymous referee for 
  very insightful suggestions in helping us to improve the paper. 
  We are grateful for research support from the Initiative
  Postdocs Support Program (No. BX201600002), 
  from the National Natural Science Foundation of China through grants
  U1631102, 11373010, 11473037, and 11633005, 
  and from the China Postdoctoral Science Foundation (grant 2017M610998). 
  This work was also supported by the National Key Research and 
  Development Program of China through grant 2017YFA0402702.}

\end{document}